\documentclass[dvips]{arxstspdf}
\usepackage{graphics}

\volume{24}
\issue{4}
\pubyear{2009}
\firstpage{398}
\lastpage{413}
\doi{10.1214/09-STS289}

\makeatletter
\newtheorem{lemma}{Lemma}[section]
\newtheorem{theorem}[lemma]{Theorem}

\def\phib{{\overline{\Phi}}}
\def\phibi{{\overline{\Phi}^{-1}}}

\makeatother

\begin{document}
\begin{frontmatter}

\title{Genome-Wide Significance Levels and Weighted Hypothesis Testing}
\runtitle{Multiple Testing}

\begin{aug}
\author[a]{\fnms{Kathryn} \snm{Roeder}\corref{}\ead[label=e1]{roeder@stat.cmu.edu}}\and
\author[a]{\fnms{Larry} \snm{Wasserman}\ead[label=e2]{wasserman@stat.cmu.edu}}
\runauthor{K. Roeder and L. Wasserman}

\affiliation{Carnegie Mellon University}

\address[a]{Kathryn Roeder is Professor of Statistics,
Department of Statistics,
Carnegie Mellon University,
5000 Forbes Avenue,
Pittsburgh, PA 15213,
USA\break \printead{e1}.
Larry Wasserman is Professor of Statistics,
Carnegie Mellon University,
5000 Forbes Avenue, Pittsburgh, PA 15213, USA\break \printead{e2}.}

\end{aug}

%
\begin{abstract}
Genetic investigations often involve the testing of vast numbers of
related hypotheses simultaneously. To control the overall error
rate, a substantial penalty is required, making it difficult to
detect signals of moderate strength. To improve the power in this
setting, a number of authors have considered using weighted $p$-values,
with the motivation often based upon the scientific plausibility of
the hypotheses. We review this literature, derive optimal weights
and show that the power is remarkably robust to misspecification of
these weights. We consider two methods for choosing weights in
practice. The first, external weighting, is based on prior
information. The second, estimated weighting, uses the data to
choose weights.
\end{abstract}

%

\end{frontmatter}

\section{Introduction}\label{sec1}

Testing for association between genetic variation and a complex
disease typically requires scanning hundreds of thousands of genetic
polymorphisms. In a multiple testing situation, such as a genome-wide
association study (GWAS), the null hypothesis is rejected for any test
that achieves a $p$-value less than a predetermined threshold (usually
on the order of $10^{-8}$). Data from these investigations has
renewed interest in the multiple testing problem. The introduction of
the false discovery rate and a procedure to control it by
Benjamini and Hochberg (\citeyear{benjaminihochberg1995}) inspired
hope that this would be an
effective way to control error while increasing power
(Storey and Tibshirani, \citeyear{storeytibshirani2003};
Sabatti, Service and Freimer, \citeyear{sabattietal2003}). To further bolster
power, recent statistical methods have been proposed that up-weight
and down-weight hypotheses, based on prior likelihood of association
with the phenotype
(Genovese, Roeder and Wasserman, \citeyear{grw2006};
Roeder et~al., \citeyear{roederetal2006};
Roeder, Wasserman and Devlin, \citeyear{roederetal2007};
Wang, Li and Bucan, \citeyear{wanglibucan2007}). Such
prior information is often available in practice.

Weighted procedures multiply the threshold by the weight $w$, for each
test, raising the threshold when $w>1$ and lowering it if $w<1$. To
control the overall rate of false positives, a budget must be imposed
on the weighting scheme, so that the average weight is one. If the
weights are informative, the procedure improves power substantially,
but, if the weights are uninformative, the loss in power is usually
small. Surprisingly, aside from this budget requirement, any set of
nonnegative weights is valid (Genovese, Roeder and Wasserman,
\citeyear{grw2006}). While
desirable in some respects, this flexibility makes it difficult to
select weights for a particular analysis.

The first such weighting scheme appears to be Holm (\citeyear
{holm1979}). Related
ideas can be found in Benjamini and Hochberg (\citeyear
{benjaminihochberg1997}),
Chen et~al. (\citeyear{chenetal2000}), Genovese, Roeder and Wasserman
(\citeyear{grw2006}), Kropf et~al. (\citeyear{kropfetal2004}),
Rosenthal and Rubin (\citeyear{rosenthalrubin1983}), Schuster, Kropf
and Roeder (\citeyear{schusteretal2004}),
Westfall and Krishen (\citeyear{westfallkrishen2001}), Westfall,
Kropf and Finos (\citeyear{westfalletal2004}),
Blanchard and Roquain (\citeyear{MR2448601}) and Roquain and van~de
Wiel (\citeyear{roquainvandewiel2008}),
among others. Several of these
approaches use data dependent weights and yet maintain familywise
error control. There are, of course, other ways to improve power
aside from weighting. Some notable recent approaches include
Rubin, Dudoit and van~der Laan (\citeyear{MR2240850}), Storey
(\citeyear{storey2007}), Donoho and Jin (\citeyear{donohojin2004}),
Signoravitch (\citeyear{signoravitch2006}), Westfall, Krishen and
Young (\citeyear{westfallkrishenyoung1998}),
Westfall and Soper (\citeyear{westfallsoper2001}), Efron (\citeyear
{efron2007}) and Sun and Cai (\citeyear{suncai2007}).
Of these, our approach is closest to Rubin, Dudoit and van~der Laan
(\citeyear{MR2240850}).

In some cases, the optimal weights can be estimated from the data. An
approach developed by Westfall, Kropf and Finos (\citeyear
{westfalletal2004}) utilizes quadratic forms
to construct such weights; however, this approach assumes the
individual measurements are normally distributed. This approach is
suited to applications such as microarray data for which the
observations are approximately normally distributed. We are
interested in applications such as tests for genetic association. In
this setting the individual observations are discrete, but the test
statistics are approximately normally distributed.

In general, $p$-value weighting raises several important questions. How much
power can we gain if we guess well in the weight assignment? How much
power can we lose if we guess poorly? In this paper we show that the
optimal weights have a simple parametric form and we investigate
various approaches for estimating these weights. We also show the
power is very robust to misspecification of the weights. In
particular, in Section~\ref{sec3} we show that (i) sparse weights (few large
weights and minimum weight close to 1) lead to huge power gains for
well specified weights, but minute power loss for poorly specified
weights; and (ii) in the nonsparse case, under weak conditions, the
worst case power for poorly specified weights is typically better than
the power obtained using equal weights.

We consider two methods for choosing the weights: (i) external
weights, where prior information (based on scientific knowledge or
prior data) singles out specific hypotheses (Section~\ref{sec4}) and (ii)
estimated weights where the data are used to construct weights
(\mbox{Section~\ref{sec5}}). External weights are prone to bias, while estimated
weights are prone to variability. The two robustness properties
reduce concerns about bias and variance.

To motivate this work consider an example (Figure~\ref{fig::linkeg})
of external weighting that arises in genetic epidemiology. To identify
variants of genes that induce greater susceptability to disease, two
types of studies (linkage and association) are often performed. Whole
genome linkage analysis has been conducted for most major diseases.
These data can be summarized by a linkage trace, a smooth stochastic
process $\{Z(s){}\dvtx{} s\in[0,L]\}$ where each $s$ corresponds to a
location on the genome. At points that correspond to a variant of a
gene of interest, the mean of the process $\mu(s)=E(Z(s))$ is a large
positive value; however, due to extensive spatial correlation in the
process, $\mu(s)$ is also nonzero in the vicinity of the variant.
Tests for association between genetic polymorphisms and disease status
for each of many genetic markers across the genome are also of
interest. Like linkage analysis, the association statistics $\{
T_j{}\dvtx{}j=1, \ldots, m\}$ map to spatial locations $\{s_j{}\dvtx
{} j=1, \ldots, m\}$
on the genome. The number of tests $m$ can be large, on the order of
1,000,000. Until recently, whole genome association analysis was
prohibitively expensive, but technological advances have now made such
studies feasible. Due to the multiple testing correction, it is
difficult to achieve sufficient power to obtain definitive results in
these studies. The linkage trace provides one obvious source of
information from which the weights can be constructed; see Section~\ref{sec6}
for further elaboration. Unlike linkage
analysis, however, the spatial correlation in association tests is
weak. For this reason, other choices such as genetic pathways could
offer a more promising source for weights in the future.

\section{Background}

\subsection{Multiple Testing}

%
\begin{figure*}

\includegraphics{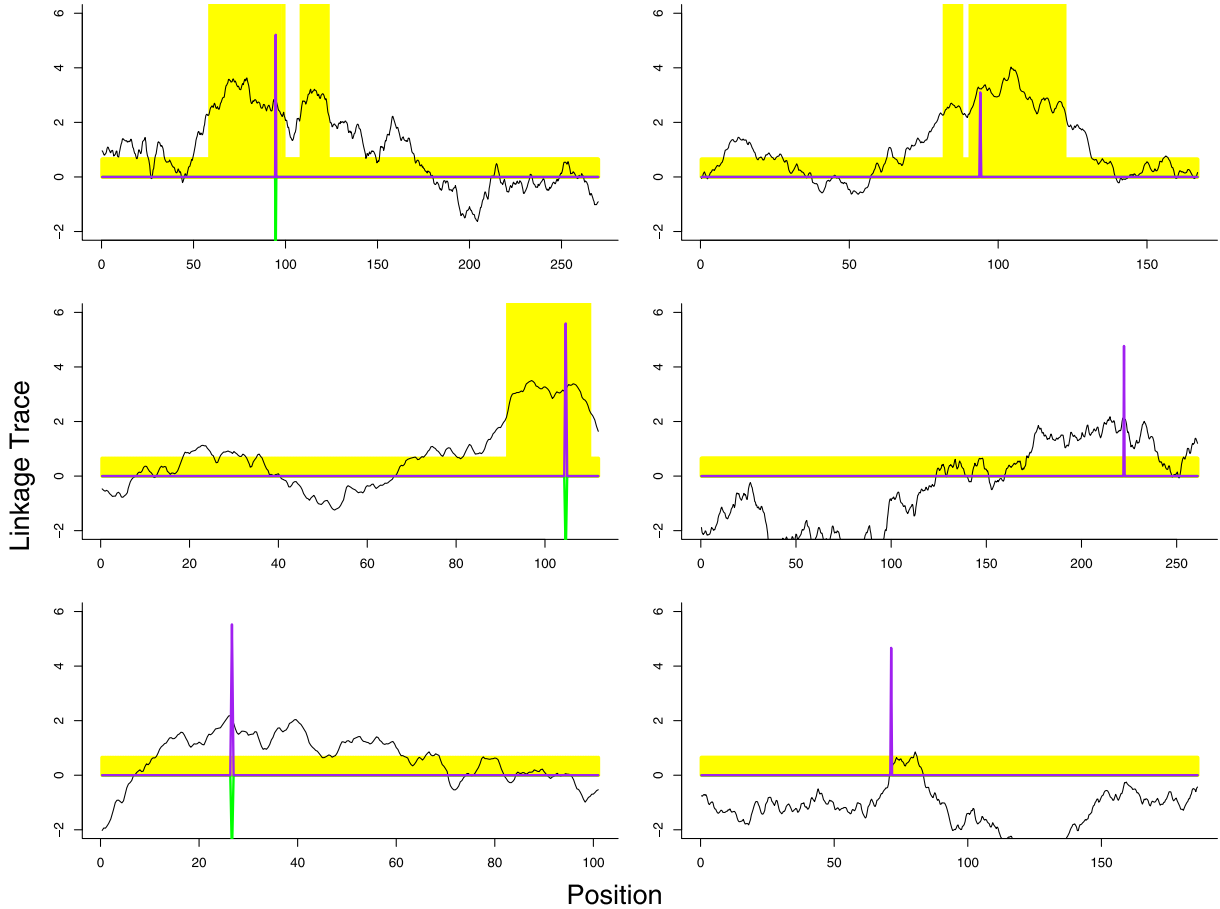}

\caption{Linkage trace and weights for 6 chromosomes. The trace is the linkage
statistic plotted as a function of position on the chromosome. The
shading indicates which $p$-values were up/down weighted. The
upspike is the association test statistic. The 3 downspikes indicate
tests that were rejected using the binary weights.}
\label{fig::linkeg}
\end{figure*}

Consider a multiple testing situation in which $m$ tests are being
performed. Suppose $m_0$ of the null hypotheses are true and $m_1 = m
- m_0$ null hypotheses are false. We can categorize the $m$ tests as
in Table~\ref{table1}. In this notation $F$ is the number of false
positives. To
control the familywise error rate, it is traditional to bound $P(F>0)$
at $\alpha$. When the tests are independent, the simplest way to
control this probability is to reject only those tests for which the
$p$-value is less than $\alpha/m$; this is called the Bonferroni
procedure.

In 1995 Benjamini and Hochberg (BH) introduced a new approach to
multiple hypothesis testing that controls the false discovery rate
(FDR), defined as the expected fraction of false rejections among those
hypotheses rejected. Let $P_{(1)} < \cdots< P_{(m)}$ be the ordered
$p$-values from $m$ hypothesis tests, with $P_{(0)} \equiv0$. Then,
the BH procedure rejects any null hypothesis for which $P \le T$ with
\[
T = \max\biggl\{ P_{(i)}{}\dvtx{} P_{(i)} \leq\frac{\alpha i}{m}
\biggr\}.
\]
This quantity is of more scientific relevance than
the overall type I error rate in GWAS.
Also, the procedure is more powerful than the Bonferroni method.
Adaptive variants of the procedure can increase power further
at little additional computational expense;
see Benjamini, Krieger and Yekutieli (\citeyear{MR2261438})
and\break Storey (\citeyear{storey2002}).

%
\begin{table}[b]
\caption{$2\times2$ classification of $m$ hypothesis tests}\label{table1}
\begin{tabular*}{\columnwidth}{@{\extracolsep{\fill}}l ccc@{}}
\hline
&$\bolds{H_0}$ \textbf{rejected} &$\bolds{H_0}$ \textbf{not
rejected}& \textbf{Total} \\
\hline
$H_0$ true & $F$ & $m_0-F$ & $m_0$ \\
$H_0$ false & $T$ & $m_1-T$ & $m_1$ \\
Total & $S$ & $m-S$ & $m$\\
\hline
\end{tabular*}
\end{table}

BH controls the false discovery rate at level $\alpha m_0/m$, where
$m_0$ is the number of true null hypotheses. With certain dependence
assumptions on the $p$-values, this is true regardless of how many nulls
are true and regardless of the distribution of the $p$-values under the
alternatives (Benjamini and Yekutieli, \citeyear{MR1869245};
Blanchard and Roquain, \citeyear{blanchardroquain2008b};
Sarkar, \citeyear{MR1892663}).
Under some distributional assumptions,
Genovese and Wasserman (\citeyear{genovesewasserman2002}) show that,
asymptotically, the BH method
corresponds to rejecting all $p$-values less than a particular $p$-value
threshold $u^*$. Specifically, $u^*$ is the solution to the equation
$H(u) =\beta u$ and $\beta=(\frac{1}{\alpha}-A_0)/(1-A_0)$, where
$A_0=m_0/m$ and $H$ is the (common) distribution of the $p$-value under
the alternative. The key result is that $\alpha/m \leq u^* \leq
\alpha$, which shows that the BH method is intermediate between
Bonferroni (corresponding to $\alpha/m$) and uncorrected testing
(corresponding to $\alpha$). If $A_0$ is close to 0, however, as it
usually is in GWA, then $\beta$ is a very large quantity and the power of
the FDR is not much improved over the Bonferroni procedure.

The power of the BH method can be improved with adaptations.
Blanchard and Roquain (\citeyear{MR2448601}) have given numerical
comparisons of
different adaptive procedures under dependence. Romano, Shaikh and Wolf
(\citeyear{MR2470085}) have considered improving the adaptive
procedure of
Benjamini, Krieger and Yekutieli (\citeyear{MR2261438}) using the
bootstrap. Sarkar and Heller (\citeyear{sarkarheller2008})
have noted that the adaptive procedure of Benjamini et al. may not
perform well compared to Storey's (\citeyear{storey2002}) procedure for certain
parameter choices.

\subsection{Weighted Multiple Testing}

We are given hypotheses $H = (H_1, \ldots, H_m)$ and standardized test
statistics $T = (T_1, \ldots, T_m)$, where $T_j \sim N(\xi_j,1)$. Likewise,
$T_j^2 \sim\chi_1^2(\xi_j^2)$. For a two-sided hypothesis, $H_j =1$ if
$\xi_j \neq0$ and $H_j =0$ otherwise. For the sake of parsimony,
unless otherwise noted, results will be stated for a one-sided test
where $H_j =1$ if $\xi_j > 0$, although the results extend easily to
the two-sided case. Let $\theta= (\xi_1, \ldots, \xi_m)$ denote the
vector of means.

The $p$-values associated with the tests are $P = (P_1, \ldots, P_m)$,
where $P_j = \overline{\Phi}(T_j)$, $\overline{\Phi} = 1- \Phi$ and
$\Phi$ denotes the standard Normal \textsc{cdf}. Let
$P_{(1)} \leq\cdots\leq P_{(m)}$
denote the sorted $p$-values and let
$T_{(1)} \geq\cdots\geq T_{(m)}$
denote the sorted test statistics.

A \textit{rejection set} $\mathcal{R}$ is a subset of $\{1, \ldots,
m\}$.
Say that $\mathcal{R}$ \textit{controls familywise error at level
$\alpha$}
if $\mathbb{P}(\mathcal{R}\cap\mathcal{H}_0)\leq\alpha$, where
$\mathcal{H}_0 = \{j{}\dvtx{}
H_j =0\}$. The \textit{Bonferroni rejection} set is
$\mathcal{R} = \{j{}\dvtx{} P_j < \alpha/m\} = \{j{}\dvtx{} T_j >
z_{\alpha/m}\}$
where we use the notation
$z_{\beta} = \overline{\Phi}^{-1}(\beta)$.

The weighted Bonferroni procedure
(Rosenthal and Rubin, \citeyear{rosenthalrubin1983};
Genovese, Roeder and Wasserman, \citeyear{grw2006})
is as follows.
Specify nonnegative weights $w=\break(w_1, \ldots, w_m)$ and
reject hypothesis $H_j$ if
%
%
\begin{equation}\label{eq::wtbonf}
j\in\mathcal{R} = \biggl\{ j{}\dvtx{} \frac{P_j}{w_j} \leq\frac
{\alpha}{m} \biggr\}.
\end{equation}
In the following lemma we show that as
long as $m^{-1}\sum_j w_j \equiv\overline w=1$, the rejection set
$\mathcal{R}$ controls familywise error at level $\alpha$. The second
lemma includes a simple modification that will be needed later.
%
\begin{lemma}\label{lemma::basic}
If $\overline w =1$,
then $\mathcal{R}$
controls familywise error at level $\alpha$.
\end{lemma}
%
%
\begin{lemma}\label{lemma::basic2}
Suppose that
$W_j = g(V_j,c)$, $j=1,\break \ldots, m$,
for some random variables $V_1, \ldots, V_m$, some constant $c$ and
some function $g$.
Further, suppose that
$V_j$ has a known distribution $H$ whenever $j\in\mathcal{H}_0$ and that
$P_j$ is
independent of $V_j$ for all $j\in\mathcal{H}_0$.
The rule that rejects when $P_j\leq\alpha W_j /m$ controls familywise
error at level $\alpha$ if $c$ is chosen to satisfy
$\mathbb{E}_H(g(V_j,c)) \leq1$.
\end{lemma}

Genovese, Roeder and Wasserman (\citeyear{grw2006}) also\break showed that
false discovery methods benefit by\break weighting.
Recall that the false discovery proportion (FDP) is
\[
{\rm FDP} = \frac{\mbox{number of false\ rejections}}{\mbox{number of rejections}} =
\frac{| \mathcal{R} \cap\mathcal{H}_0 |}{| \mathcal{R}|},
\]
where the ratio is defined to be 0 if the denominator is~0.
The false discovery rate (FDR) is
$\mathrm{FDR} = \mathbb{E}({\rm FDP})$.
Benjamini and Hochberg (\citeyear{benjaminihochberg1995}) proved\break
$\mathrm{FDR} \leq\alpha$ if
$\mathcal{R} = \{ j{}\dvtx{} P_{(j)} \leq T\}$
where
$T =\break \max\{ j{}\dvtx{} P_{(j)} \leq j\alpha/m\}$.
Genovese, Roeder and Wasserman (\citeyear{grw2006}) showed that
$\mathrm{FDR} \leq
\alpha$ if the $P_j$'s are replaced by $Q_j = P_j/w_j$ provided
$\overline w =1$.
This paper focuses on familywise error
using the weighted procedure (\ref{eq::wtbonf}).
Similar results hold for FDR
and other familywise
controlling procedures such as Holm's test.

\section{Power, Robustness and Optimality}\label{sec3}

The optimal weights, derived below, can be re-expressed as optimal
cutoffs for testing. Specifically, rejecting when $P_j/w_j \leq
\alpha/m$ is the same as rejection when $T_j > \xi_j/2 + c/\xi_j$.
This result can be obtained from Spj{\o}tvoll (\citeyear
{spjotvoll1972}) and
is identical to the result in Rubin, Dudoit and van~der Laan (\citeyear
{MR2240850}) obtained
independently. The remainder of the paper, which shows some good
properties of the weighted method, can thus also be considered as
providing support for their method for selecting test specific
cutoffs. In particular, Rubin et al. (\citeyear{MR2240850})'s simulations indicate
that even poorly specified estimates of the cutoffs $\xi_j/2 +
c/\xi_j$ can still perform well. In this section we provide insight
into why this is true.

The power of a single, one-sided alternative in the unweighted case
($w_j=1$) is
\[
\pi(\xi_j,1) = \mathbb{P}(T_j > z_{\alpha/m})
= \overline{\Phi}(z_{\alpha/m} - \xi_j).
\]
The power\footnote{For a two-sided alternative the power is
\[
\pi(\xi_j,w_j)= \overline{\Phi}(z_{\alpha w_j/2m}-\xi_j )
+\overline{\Phi}(z_{\alpha w_j/2m}+\xi_j ).\vspace*{-12pt}
\]
}
in the weighted case is
%
%
\begin{eqnarray}\label{eq::power-formula}
\pi(\xi_j,w_j)
&=&\mathbb{P}\biggl( P_j < \frac{\alpha w_j}{m} \biggr)\nonumber
\\
&=& \mathbb{P}\biggl( T_j > \overline{\Phi}^{-1} \biggl(\frac{\alpha w_j}{m} \biggr) \biggr)
\\
&=&\overline{\Phi}( z_{\alpha w_j/m}-\xi_j ).\nonumber
\end{eqnarray}
Weighting increases the power when
$w_j >1$ and decreases the power when $w_j < 1$ for the $j$th alternative.

Given $\theta= (\xi_1, \ldots, \xi_m)$ and
$w = (w_1, \ldots, w_m)$, we define the \textit{average power}
\[
\frac{1}{m_1}\sum_{j=1}^m \pi(\xi_j,w_j) I(\xi_j > 0),
\]
where $m_1=\sum_{j=1}^m I(\xi_j > 0)$.
More generally, if $\xi$ is drawn from a distribution $Q$
and $w = w(\xi)$ is a weight function,
we define the average power
$\int\pi(\xi,\break w(\xi))I(\xi>0) \,dQ(\xi)/\int I(\xi>0) \,dQ(\xi)$.
If we take $Q$ to be the empirical distribution of
$(\xi_1, \ldots, \xi_m)$,
then this reduces to the previous expression.
In this formulation we require $w(\xi) \geq0$ and $\int w(\xi)\,
dQ(\xi) =1$.

%
\begin{figure}[b]

\includegraphics{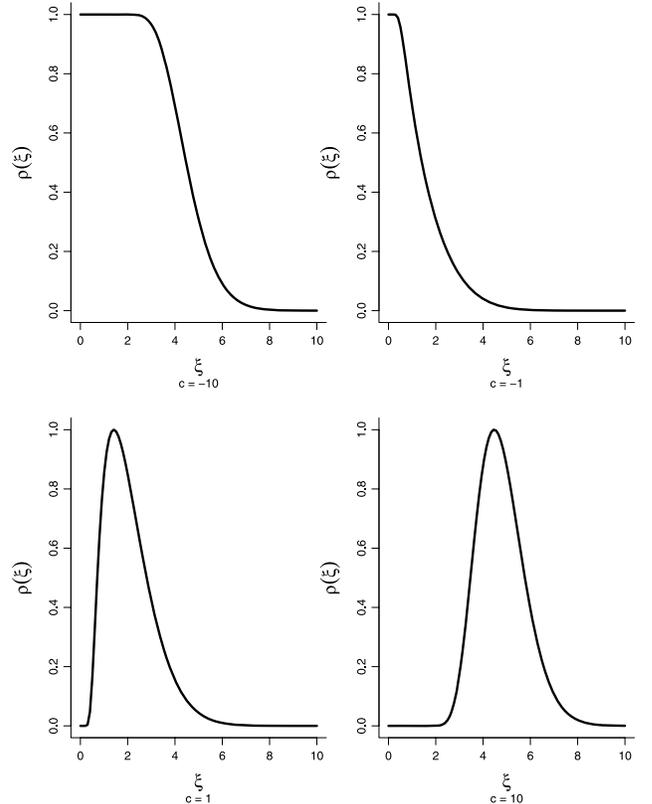}

\caption{Optimal weight function $\rho_c(\xi)$ for various $c$.
In each case $m=1000$ and $\alpha= 0.05$.
The functions are normalized to have maximum 1.}\label{fig::family}
\end{figure}

In the following theorem
we see that the set of optimal weight functions form a one parameter
family indexed by a constant $c$.
%
\begin{theorem}\label{thm::optimal}
Given
$\theta=(\xi_1, \ldots, \xi_m)$,
the optimal weight
vector $w=(w_1, \ldots, w_m)$ that maximizes the average power
subject to $w_j \geq0$ and $\overline w=1$
is $w=(\rho_c(\xi_1),\ldots,\rho_c(\xi_m))$,
where
%
%
\begin{equation}\label{eq:wq1}
\rho_c(\xi) =
\biggl(\frac{m}{\alpha} \biggr)\overline{\Phi} \biggl(\frac{\xi
}{2} + \frac{c}{\xi} \biggr)I(\xi> 0),
\end{equation}
and $c\equiv c(\theta)$ is defined by the condition
%
%
\begin{equation}
\label{eq:wbest}
\frac{1}{m} \sum_{j=1}^m \rho_c(\xi_j) =1.
\end{equation}
\end{theorem}

The proof, essentially a special case of Spj{\o}tvoll (\citeyear
{spjotvoll1972}),
is in the \hyperref[appendix]{Appendix}. Figure~\ref{fig::family} displays the
function $\rho_c(\xi)$ for various values of $c$ (the function is
normalized to have maximum 1 for easier visualization). The result
generalizes to the case where the alternative means are random
variables with distribution $Q$ in which case $c$ is defined by $\int\rho_c(\xi)\,dQ(\xi) =1$.

From (\ref{eq::power-formula}) and (\ref{eq:wq1}) we have immediately:
\begin{lemma}\label{lemma::power-at-opt}
The power at an alternative with mean $\xi$ under
optimal weights is
$\overline{\Phi} (c/\xi- \xi/2 )$.
The average power under optimal weights, which we call the
\textup{oracle power}, is
\[
\frac{1}{m_1}\sum_{j=1}^m \overline{\Phi}
\biggl(\frac{c}{\xi_j} - \frac{\xi_j}{2} \biggr)I(\xi_j > 0),
\]
where $m_1 = \sum_j I(\xi_j > 0)$.
\end{lemma}

The oracle power is not attainable since
the optimal weights depend on
$\theta=(\xi_1, \ldots,\xi_m)$.
In practice, the weights will either be chosen by
prior information or by estimating the $\xi$'s.
This raises the following question:
how sensitive is the power to correct specification of
the weights?
Now we show that the power is very robust to weight misspecification.
\begin{longlist}
\item[\textit{Property} I:] Sparse weights (minimum weight close to
1) are
highly robust. If most weights are less than~1 and the minimum weight
is close to 1, then correct specification (large weights on
alternatives) leads to large power gains but incorrect specification
(large weights on nulls) leads to little power loss.

\item[\textit{Property} II:]
Worst case analysis.
Weighted hypothesis testing,
even with poorly chosen weights,
typically does as well or better than
Bonferroni except when the
the alternative means are large,
in which both have high power.
\end{longlist}

Let us now make these statements precise.
Also, see Genovese, Roeder and Wasserman (\citeyear{grw2006}) and
Roeder et~al. (\citeyear{roederetal2006}) for other results on the
effect of weight misspecification.

\textit{Property} I.
Consider first the case where the weights take two distinct values
and the alternatives have a common mean $\xi$.
Let $\varepsilon$ denote the
fraction of hypotheses given the
larger of the two values of the weights $B$.
Then, the weight vector $w$ is proportional to
\[
(\underbrace{B, \ldots, B}_{k \ \mathrm{terms}},
\underbrace{1, \ldots, 1}_{m-k \ \mathrm{terms}}),
\]
where $k = \varepsilon m$ and $B > 1,$
and, hence, the normalized weights are
\[
w = (\underbrace{w_1, \ldots, w_1}_{k \ \mathrm{terms}},
\underbrace{w_0, \ldots, w_0}_{m-k \ \mathrm{terms}}),
\]
where
\[
w_1 = \frac{B}{ \varepsilon B + (1-\varepsilon)},\quad
w_0 = \frac{1}{ \varepsilon B + (1-\varepsilon)}.
\]
We say that the weights are sparse if $\varepsilon$ is small. Provided
$B$ is considerably less than $1/\varepsilon$,
most weights are near 1 in the sparse case.

\begin{figure}[b]

\includegraphics{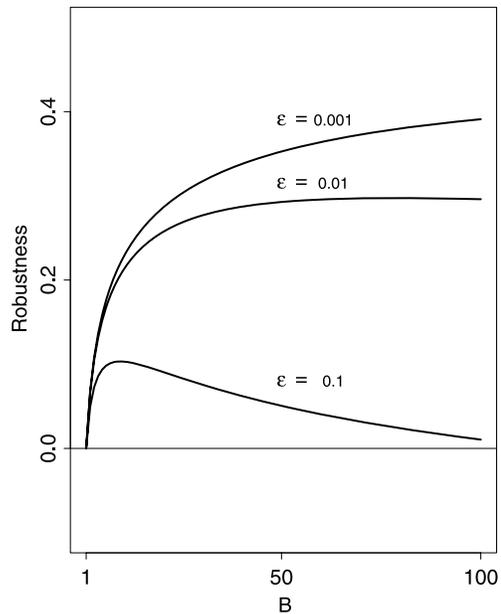}

\caption{Robustness function for $m=1000$.
In this example, $\xi= z_{\alpha/m}$ which has
power 1$/$2 without weighting.
The gain of correct weighting far outweighs the loss for
incorrect weighting as long as the fraction of large weights $\varepsilon$
is small.}\label{fig::powermult}
\end{figure}

Rather than investigate the average power, we focus on
a single alternative with mean $\xi$.
The power gain by up-weighting this hypothesis
is the power under weight $w_1$ minus the unweighted power\break
$\pi(\xi,w_1)-\pi(\xi,1)$.
Similarly, the power loss for down-weighting is
$\pi(\xi,1)-\pi(\xi,w_0)$.
The gain minus the loss,
which we call the robustness function, is
\begin{eqnarray*}
R(B,\varepsilon) & \equiv&
\bigl(\pi(\xi,w_1) - \pi(\xi,1) \bigr)
\\
&&{}+ \bigl(\pi(\xi,1) - \pi(\xi,w_0) \bigr)\\
&=&
\overline{\Phi} ( z_{\alpha w_1/m} - \xi) +
\overline{\Phi} ( z_{\alpha w_0/m} - \xi)
\\
&&{}-2\overline{\Phi} ( z_{\alpha/m} - \xi).
\end{eqnarray*}
The gain outweighs the loss if and only if $R(B,\varepsilon) >0$
(Figure~\ref{fig::powermult}).

%

In the sparse weighting scenario $k$ is small and $w_0\approx1$ by
assumption, consequently, an analysis of $R(B,\varepsilon)$ sheds light
on the effect of weighting on power, without the added complications
involved in a full analysis of average power.
\begin{theorem}
Fix $B>1$.
Then,
$\lim_{\varepsilon\to0}R(B,\break\varepsilon) > 0$.
Moreover,
there exists $\varepsilon^*(B)>0$ such that
$R(B,\varepsilon)>0$ for all $\varepsilon< \varepsilon^*(B)$.
\end{theorem}

We can generalize this beyond the two-valued case as follows.
Let $w$ be any weight vector such that $\overline w=1$.
Now define the (worst case) robustness function
\begin{eqnarray*}
R(\xi) &\equiv&\min_{\{j: w_j > 1, H_j=1\}}
\{ \pi(\xi,w_j) - \pi(\xi,1) \}
\\
&&{}-\max_{\{j: w_j < 1, H_j=1\}}\{ \pi(\xi,1) - \pi(\xi,w_j) \}.
\end{eqnarray*}
We will see that $R(\xi) >0$ under weak conditions and that
the maximal robustness is obtained for $\xi$ near
the Bonferroni cutoff $z_{\alpha/m}$.
\begin{theorem}\label{thm::sparse-is-good}
A necessary and sufficient condition for $R(\xi) >0$ is
%
%
\begin{eqnarray}\label{eq::cond-for-robust}
R_{b,B}(\xi) &\equiv&{\Phi} ( z_{\alpha B/m}-\xi)
+{\Phi} ( z_{\alpha b/m}-\xi)\nonumber
\\[-8pt]\\[-8pt]
&&{}-2{\Phi} ( z_{\alpha/m}-\xi)\leq0,\nonumber
\end{eqnarray}
where
$B= \min\{w_j: w_j > 1\}$,
$b = \min\{w_j\}$.
Moreover,
\[
R_{b,B}(\xi) = -\Delta(\xi) + O(1-b),
\]
where
\[
\Delta(\xi) = \bigl({\Phi} ( z_{\alpha/m}-\xi)-{\Phi} ( z_{\alpha B/m}-\xi) \bigr) > 0
\]
and, as $b\to1$,
$\mu(\{\xi\dvtx R(\xi) < 0\})\to0$ and\break
$\inf_\xi R(\xi) \to0$.
\end{theorem}

Based on the theorem, we see that there is overwhelming robustness as
long as the minimum weight is near 1. Even in the extreme case $b=0$,
there is still a safe zone, an interval of values of $\xi$ over which
$R(\xi)>0$.
\begin{lemma}\label{lemma::safe}
Suppose that $B\geq2$.
Then there exists $\xi_* > 0$ such that
$R_{B,b}(\xi) >0$ for all $0\leq\xi\leq\xi_*$ and all $b$.
An upper bound on $\xi_*$ is
$z_{\alpha/m} - 1/(z_{\alpha/m} - z_{B \alpha/m})$.
\end{lemma}

\textit{Property} II. Even if the weights are not sparse, the power of
the weighted test tends to be acceptable.

The result holds even
though the weights themselves can be very sensitive to changes in
$\theta$. Consider the following example. Suppose that
$\theta=(\xi_1, \ldots, \xi_m)$ where each $\xi$ is equal to
either 0
or some fixed number $\xi$. The empirical distribution of the
$\xi_j$'s is thus $Q=(1-a)\delta_0 + a \delta_\xi$, where $\delta$
denotes a point mass and $a$ is the fraction of nonzero means. The
optimal weights are $0$ for $\xi_j=0$ and $1/a$ for $\xi_j=\xi$. Let
$\tilde{Q}=(1-a-\gamma)\delta_0 + \gamma\delta_u + a \delta_\xi
$, where
$u$ is a small positive number. Since we have only moved the mass at
0 to $u$, and $u$ is small, we would hope that $w(\xi)$ will not
change much. But this is not the case. Set $\xi= A + \sqrt{A2 -
2c}$, $u = B - \sqrt{B2 - 2c}$, where
\begin{eqnarray*}
A &=& \overline{\Phi}^{-1} \biggl(\frac{\alpha}{(m(\gamma K+a))}
\biggr),\\
B &=& \overline{\Phi}^{-1} \biggl(\frac{K \alpha}{(m(\gamma K+a))} \biggr).
\end{eqnarray*}
This arrangement yields weights $w_0$ and $w_1$ on $u$ and $\xi$ such that
$w_0/w_1=K$.
For example, if
$m= 1000$, $\alpha= 0.05$,
$a =0.1$,
$\gamma=0.1$,
$K = 1000$
and $c =0.1$, then $u=0.03$ and $\xi= 9.8$.
The optimal weight on $\xi$ under $Q$ is
10 but under $\tilde{Q}$ it is
$0.00999$ and so is reduced by a factor of 1001.
More generally, we have the following result
which shows that the weights are,
in a certain sense, a discontinuous function of $\theta$.
\begin{lemma}\label{lemma::discon}
Fix $\alpha$ and $m$.
For any $\delta>0$ and $\varepsilon>0$
there exists
$Q=(1-a)\delta_0 + a\delta_\xi$ and
$\tilde{Q}=(1-a-\gamma)\delta_0 +\gamma\delta_u + a\delta_\xi$
such that
$d(Q,\tilde{Q}) < \delta$, and\break
$\tilde\rho(\xi)/\rho(\xi) < \varepsilon$,
where
$a=\alpha/4$,
$d(Q,\tilde{Q}) =\break \sup_{\xi} | Q(-\infty, \xi], \tilde{Q}(-\infty
, \xi]|$
is the Kolmogorov--\break Smnirnov distance,
$\rho$ is the optimal weight function for $Q$ and
$\tilde{\rho}$ is the optimal weight function for $\tilde{Q}$.
\end{lemma}

\begin{figure*}

\includegraphics{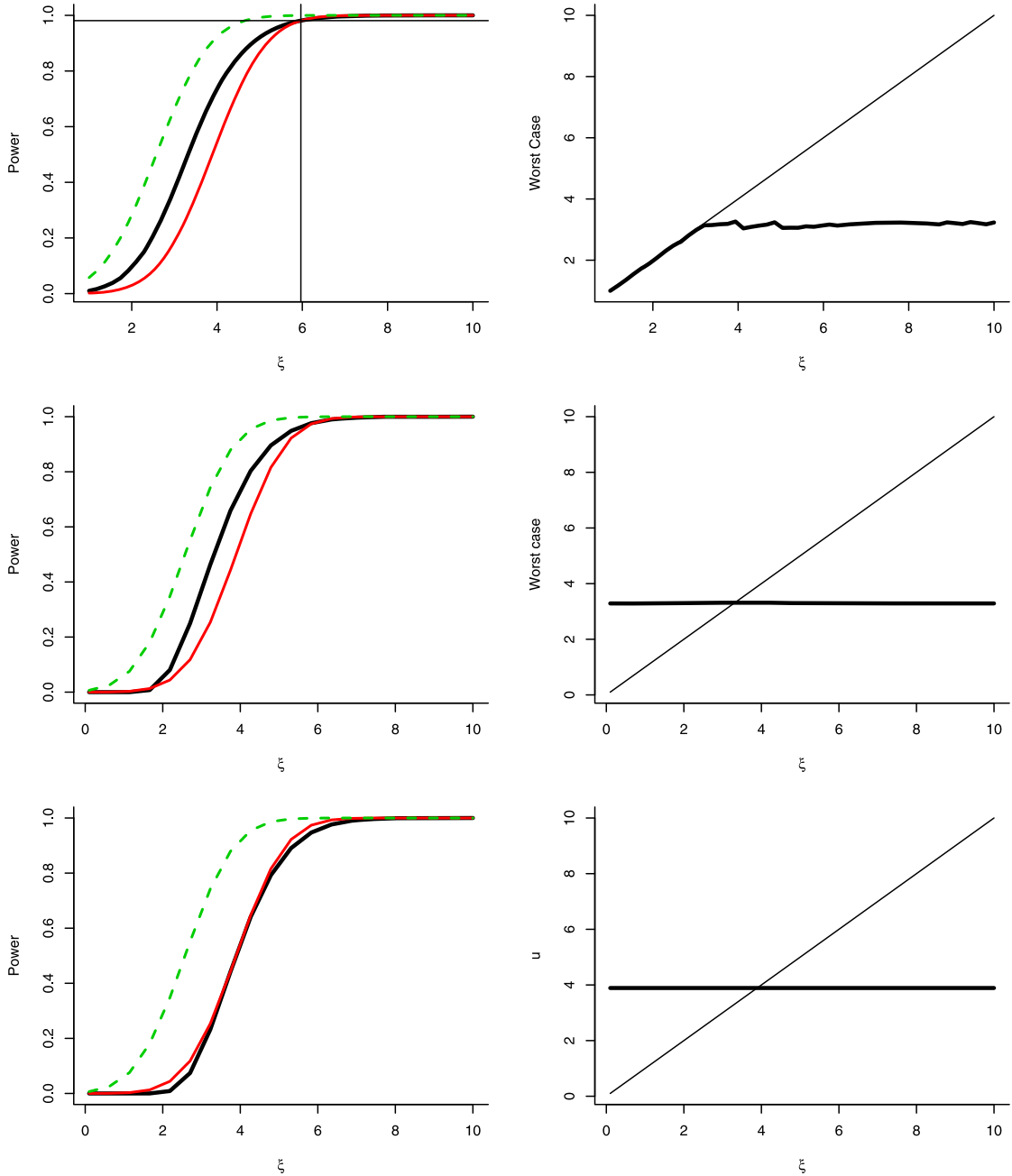}

\caption{Power as a function of the alternative mean $\xi$.
In these plots, $a=0.01$, $m=1000$ and $\alpha=0.05$.
There are $(1-a)m$ nulls and $ma$ alternatives with mean $\xi$.
The left plots show what happens when the weights are
incorrectly computed assuming that a fraction $\gamma$ of nulls
are actually alternatives with mean $u$.
In the top plot, we restrict $0 < u < \xi$.
In the second and third plots, no restriction is placed on $u$.
The top and middle plots have $\gamma= 0.1,$ while
the third plot has $\gamma=1-a$ (all nulls misspecified as alternatives).
The dark solid line shows the lowest possible power assuming
the weights were estimated as poorly as possible.
The lighter solid line is the power of the unweighted (Bonferroni) method.
The dotted line is the power under the optimal weights.
The vertical line in the top plot is at $\xi_*$. The weighted method
beats unweighted for all $\xi< \xi_*$.
The right plot shows the least favorable $u$ as a function of $\xi$.
That is, mistaking $\gamma m$ nulls for alternatives with mean $u$
leads to the worst power.
Also shown is the line $u=\xi$.}
\label{fig::minimax}
\end{figure*}

Fortunately, this feature of the weight function does not pose
a serious hurdle in practice because
it is possible to have high power even with poor weights.
In Figure~\ref{fig::minimax} the plots
on the left show the power as a function of the alternative mean
$\xi$. The dark solid line shows the lowest possible power assuming
the weights were estimated as poorly as possible (under conditions
specified below). The lighter solid line is the power of the
unweighted (Bonferroni) method. The dotted line shows the power under
theoretically optimal weights. The worst case weighted power is
typically close to or larger than the Bonferroni power except for
large $\xi$ when they are both large.

%

To begin formal analysis, assume that each mean is
either equal to $0$ or $\xi$ for some fixed $\xi>0$. Thus, the
empirical distribution is $Q = (1-a)\delta_0 + a \delta_\xi$, where
$\delta$ denotes a point mass and $a$ is the fraction of nonzero
$\xi_j$'s. The optimal weights are $1/a$ for hypotheses whose mean is
$\xi$. To study the effect of misspecification error, consider the
case where $\gamma m$ nulls are mistaken for alternatives with mean
$u>0$. This corresponds to misspecifying $Q$ to be $\tilde{Q} =
(1-a-\gamma)\delta_0 + \gamma\delta_u + a \delta_\xi$. We will study
the effect of varying $u$, so let $\pi(u)$ denote the power at the true
alternative $\xi$ as a function of $u$. Also, let $\pi_{\rm Bonf}$
denote the power using equal weights (Bonferroni). Note that changing
$Q = (1-a)\delta_0 + a \delta_\xi$ to $Q = (1-a)\delta_0 + a
\delta_{\xi'}$ for $\xi' \neq\xi$ does not change the weights.

As the weights are a function of $c$, we first need to find
$c$ as a function of $u$.
The normalization condition~(\ref{eq:wbest}) reduces to
%
%
\begin{equation}\label{eq::wbest2}
\gamma\overline{\Phi} \biggl(\frac{u}{2} + \frac{c}{u} \biggr) +
a \overline{\Phi} \biggl(\frac{\xi}{2} + \frac{c}{\xi} \biggr)
=\frac{\alpha}{m},
\end{equation}
which implicitly defines the function $c(u)$.
First we consider what happens when $u$ is restricted to be less than
$\xi$.
\begin{theorem}\label{thm::restrict-u}
Assume that $\alpha/m \leq\gamma+ a \leq1$.
Let $Q = (1-a)\delta_0 + a \delta_\xi$ and
$\tilde{Q} = (1-a-\gamma)\delta_0 + \gamma\delta_u + a \delta_\xi$
with $0\leq u \leq\xi$.
Let $C(\xi) = \sup_{0\leq u \leq\xi}c(u)$ and define
$\xi_0 = z_{\alpha/(m(\gamma+a))}$:
\begin{enumerate}
\item For $\xi\leq\xi_0$, $C(\xi) = \xi\xi_0- \xi^2/2$.
For $\xi> \xi_0$, $C(\xi)$ is the solution to
\[
\gamma\overline{\Phi}\bigl(\sqrt{2c}\bigr) + a\overline{\Phi} \biggl(\frac{c}{\xi} + \frac{\xi}{2} \biggr) =
\frac{\alpha}{m}.
\]
In this case,
$C(\xi) = z^2_{\alpha/(m\gamma)}/2 + O(a)$.
\item Let $\xi_* =
z_{\alpha/m} +
\sqrt{z_{\alpha/m}^2 - z_q^2}$, where $q = \alpha(1-a)/(m\gamma)$.
For $\xi< \xi_*$,
%
%
\begin{equation}
\label{eq::thebonfclaim}
\inf_{0 < u < \xi}\pi(u) \geq\pi_{\rm Bonf}.
\end{equation}
For $\xi\geq\xi_*$ we have
%
%
\begin{eqnarray}
\label{eq::thenextclaim}
\hspace*{25pt}\inf_{0 < u < \xi}\pi(u) &\geq&
\overline{\Phi} \biggl( \frac{ z^2_{\alpha/(m\gamma)} - \xi_*^2}{2\xi_*} \biggr)- O(a)\\
&\approx& 1-\overline{\Phi} \Biggl(\sqrt{2\log\frac{1-a}{\gamma}}\Biggr)-O(a)\\
& \geq& 1- \frac{\gamma}{1-a} - O(a).
\end{eqnarray}
\end{enumerate}
\end{theorem}

The factor
$\overline{\Phi} (\sqrt{2\log(1-a)/\gamma} )\approx\gamma/(1-a)$
is the worst case power deficit due to misspecification.
Now we drop the assumption that $u\leq\xi$.
\begin{theorem}\label{thm::drop-u}
Let $Q=(1-a)\delta_0 + a\delta_\xi$ and
let
$Q_u \equiv(1-a-\gamma)\delta_0 + \gamma\delta_u +a\delta_\xi$.
Let $\pi_u$ denote the power at $\xi$ using the weights
computed under $Q_u$.
\begin{enumerate}
\item
The least favorable $u$ is
$u_* \equiv\mathop{\arg\min}_{u\geq0} \pi_u =\break \sqrt{2c_*} = z_{\alpha/(m\gamma)} + O(a)$,
where
$c_*$ solves
\[
\gamma\overline{\Phi}\bigl(\sqrt{2 c_*}\bigr) + a \overline{\Phi} \biggl( \frac{\xi}{2}+\frac{c_*}{\xi} \biggr)=\frac{\alpha}{m}
\]
and
$c_* = z^2_{\alpha/(m\gamma)}/2 + O(a)$.
\item The minimal power is
\[
\inf_u \pi_u = \overline{\Phi} \biggl(\frac{c_*}{\xi} - \frac{\xi}{2}\biggr)
=\overline{\Phi} \biggl(\frac{z_{\alpha/(m\gamma)}^2-\xi^2}{2\xi} \biggr) + O(a).
\]
\item A sufficient condition for
$\inf_u \pi_u$ to be larger than the power of the Bonferroni method is
$\xi\geq z_{\alpha/m} + \sqrt{z^2_{\alpha/m} - z^2_{\alpha
/(m\gamma)}} + O(a)$.
\end{enumerate}
\end{theorem}
%

\section{Choosing External Weights}\label{sec4}

%
\begin{figure}[b]

\includegraphics{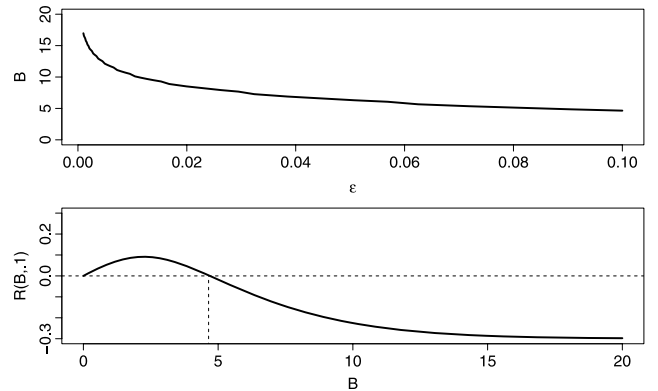}

\caption{Top plot: turnaround point $B_0(\varepsilon)$ versus $\varepsilon$.
Bottom plot shows the robustness function $R(B,0.1)$
versus $B$. The turnaround point $B_0(\varepsilon)$ is shown
with a vertical dotted line.}\label{fig::turnaround}
\end{figure}

One approach to choosing external weights
(or test statistic cutoffs)
is to use empirical Bayes methods to model prior information
while being careful to preserve error control
as in Westfall and Soper (\citeyear{westfallsoper2001}), for example.
Here we consider a simple\break method that takes advantage of the robustness
properties we have discussed.
We will focus here on the two-valued case.
Thus,
\[
w = (\underbrace{w_1, \ldots, w_1}_{k \ \mathrm{terms}},
\underbrace{w_0, \ldots, w_0}_{m-k \ \mathrm{terms}}),
\]
where $k=\varepsilon m$,
$w_1 = B/( \varepsilon B + (1-\varepsilon))$ and
$w_0 = 1/( \varepsilon B + (1-\varepsilon))$.
In practice, we would typically have a fixed fraction of hypotheses
$\varepsilon$
that we want to give more weight to.
The question is how to choose $B$.
We will focus on choosing $B$ to produce weights with good properties
at interesting values of $\xi$.
Now large values of $\xi$ already have high power.
Very small values of $\xi$ have extremely low power and benefit little
by weighting.
This leads us to focus on
constructing weights that are
useful for a \textit{marginal effect}, defined as the alternative $\xi_0$
that has power 1$/$2 when given weight 1.
Thus, the marginal effect is $\xi_0 = z_{\alpha/m}$.
In the rest of this section then
we assume that all nonzero $\xi_j$'s are equal to $\xi_0$.
Of course, the validity of the procedure does not depend on this
assumption being true.

Fix $0 < \varepsilon< 1$ and vary $B$.
As we increase $B$, we will eventually reach a point
$B_0(\varepsilon)$ where
$R(B,\varepsilon) < 0$, which we call
the turnaround point.
Formally, $B_0(\varepsilon) = \sup\{ B{}\dvtx{} R(B,\varepsilon) > 0 \}$.
The top panel in Figure~\ref{fig::turnaround}
shows
$B_0(\varepsilon)$ versus $\varepsilon$,
which shows that for small $\varepsilon$ we can choose $B$ large
without loss of power.
The bottom panel shows
$R(B,\varepsilon)$ for $\varepsilon= 0.1$.
Ideally, for a given $\varepsilon,$ one chooses
$B$ near $B_*(\varepsilon)$, the value of $B$ that maximizes
$R(B,\varepsilon)$.
\begin{theorem}
Fix $0 < \varepsilon< 1$.
As a function of $B$,
$R(B,\varepsilon)$ is unimodal and satisfies
$R(1,\varepsilon) =1$,
$R'(1,\varepsilon) >0$ and
$R(\infty,\varepsilon) < 0$.
Hence,
$B_0(\varepsilon)$ exists and is unique.
Also, $R(B,\varepsilon)$ has a unique maximum at some point
$B^*(\varepsilon)$ and
$R(B^*(\varepsilon),\varepsilon)>0$.
\end{theorem}

When $\varepsilon$ is very small, $B$ can be large, provided
$w_0\approx1$.
For example, suppose we want to increase the chance of
rejecting one particular hypothesis so that $\varepsilon= 1/m$. Then,
\[
w_1 = \frac{mB}{B+m -1}\approx B,\quad w_0 = \frac{1}{B+m -1}\approx1
\]
and
\begin{eqnarray}
\lim_{m\to\infty}\lim_{B\to\infty} \pi(\xi_j,w_1) =1,\nonumber
\\
\eqntext{\displaystyle\mbox{while} \lim_{m\to\infty}\lim_{B\to\infty} \pi(\xi_j,w_0)=\frac{1}{2}.}
\end{eqnarray}

The next results show that binary weighting schemes
are optimal in a certain sense.
Suppose we want to have at least a fraction $\varepsilon$
with high power $1-\beta$ and otherwise we want to
maximize the minimum power.
\begin{theorem}
\label{thm::method1}
Consider the following optimization problem:
Given $0 < \varepsilon< 1$ and
$0 < \beta< 1/2$, find a vector
$w = (w_1, \ldots, w_m)$ that maximizes
$\min_j \pi(\xi_m,\break w_j)$
subject\vspace*{1pt}
$\overline{w} =1$, and
$\# \{ j{}\dvtx{} \pi(w_j,\xi_m)\geq1-\beta\}/m \geq\varepsilon$.
The solution is given by
$c = \overline{\Phi} ( z_{\alpha/m} + z_{1-\beta} )$,
$B = cm(1-\varepsilon)/(\alpha- \varepsilon c m)$,
$w_1 = B/( \varepsilon B + (1-\varepsilon))$,
$w_0 = 1/( \varepsilon B + (1-\varepsilon))$ and
$k = \varepsilon m$.
\end{theorem}

If our goal is to maximize the number of alternatives with
high power while maintaining a minimum power loss,
the solution is given as follows.
\begin{theorem}
\label{thm::method2}
Consider the following optimization problem:
Given $0 < \beta< 1/2$, find a vector
$w = (w_1, \ldots, w_m)$ that maximizes
$\# \{ j{}\dvtx{} \pi(w_j,\xi_m)\geq1-\beta\}$
subject to
$\overline{w} =1,  \mbox{ and }
\min_j \pi(w_j,\xi_m) \geq\delta$.
The solution is
\begin{eqnarray*}
w_1 &=& \frac{m}{\alpha}\overline{\Phi} (z_{\alpha/m} + z_{1-\beta}), \quad
w_0 = \frac{m}{\alpha}\overline{\Phi} (z_{\alpha/m} + z_{\delta}),
\\
\varepsilon&=& \frac{1-w_0}{w_1-w_0}
\end{eqnarray*}
and $k = m \varepsilon$.
\end{theorem}

A special case that falls under this theorem permits the minimum power
to be 0. In this case $w_0=0$ and $\varepsilon=1/w_1$.

\section{Estimated Weights}\label{sec5}

In practice, $\xi_j$ is not known, so it must be estimated to utilize the
weight function. A natural choice is to build on the two stage
experimental design
(Satagopan and Elston, \citeyear{satagopanelston2003};
Wang et~al., \citeyear{wangetal2006})
and split the data into subsets, using one subset to estimate $\xi_i$,
and hence $w(\xi_i)$, and the second to conduct a weighted test of the
hypothesis (Rubin, Dudoit and van~der Laan, \citeyear{MR2240850}).
This approach would arise naturally
in an association test conducted in stages. It does lead to a gain in
power relative to unweighted testing of stage 2 data; however, it is
not better than simply using the full data set without weights for the
analysis (Rubin, Dudoit and van~der Laan, \citeyear{MR2240850}). These results
are corroborated by Skol et~al. (\citeyear{skoletal2006}) in a
related context. They
showed that it is better to use stages 1 and 2 jointly, rather than
using stage 2 as an independent replication of stage 1.

To gain a strong advantage with data-based weights, prior information
is needed. One option is to order the tests (Rubin, Dudoit and van~der
Laan, \citeyear{MR2240850}), but
with a large number of tests this can be challenging.
The type of prior information readily available to investigators is
often nonspecific. For instance, SNPs might naturally be grouped,
based on features that make various candidates more promising for this
disease under investigation. For a brain-disorder phenotype we might
cross-classify SNPs by categorical variables such as functionality,
brain expression and so forth. The SNPs in one group may seem most
promising, a priori, while those in another seem least promising.
Intermediate groups may be somewhat ambiguous. It is easy to imagine
additional variables that further partition the SNPs into various
classes that help to separate the more promising SNPs from the others.
While this type of information lends itself to grouping SNPs, it does
not lead directly to weights for the groups. Indeed, it might not even
be possible to choose a natural ordering of the groups. What is
needed is a way to use the data to determine the weights, once the
groups are formed.

Until recently, methods for weighted multiple-testing
required that prior weights be developed independently of the data
under investigation
(Genovese, Roeder and Wasserman, \citeyear{grw2006};
Roeder, Wasserman and Devlin, \citeyear{roederetal2007}). Here
we provide a data-based estimate of weights based on results of grouped
analysis. One way to implement this approach is to follow these steps:
\begin{longlist}[1.]
\item[1.] Partition the tests into subsets $\mathcal{G}_1,\ldots,\mathcal
{G}_K$, with the
$k$th group containing $r_k$ elements, ensuring that $r_k$ is at
least 20--30.

\item[2.] Calculate the sample mean $Y_k$ and variance $S_k^2$ for the
test statistics in each group.

\item[3.] Label the $i$th test in group $k$, $T_{ik}$. At best, only a
fraction of the elements in each group will have a signal, hence, we
assume that for $i=1,\ldots,r_k$ the distribution of the test
statistics is approximated by a mixture model
\[
T_{ik} \sim(1-\pi_k) N(0,1) + \pi_k N(\xi_k,1)
\]
or
\[
T_{ik} \sim(1-\pi_k) \chi_1^2(0) + \pi_k \chi_1^2(\xi_k^2),
\]
where $\xi_{k}$ is the signal size for those tests with a signal in
the $k$th group. This is an approximation because the signal is
likely to vary across tests. The mixture of normals is only
appropriate when the tests are one-sided. For two-sided alternatives,
the $\chi^2$ is the natural approach. This test squares the
noncentrality parameter, effectively removing any ambiguity about the
direction of the associations.

\item[4.] Estimate $(\pi_k,\xi_k)$ using the method of moments estimator
(for details see the \hyperref[appendix]{Appendix}). Because $\xi_k$ has no meaning when
$\pi_k =0$, the $\hat\xi_k$ is set to 0 when $\hat\pi_k$ is close
to zero. For the normal model the estimators are
%
%
\begin{equation}
\label{eq:norm}
\hat\pi_k= Y_k^2/(Y_k^2 + S_k^2 - 1),\quad
\hat\xi_k = Y_k/\pi_k,
\end{equation}
provided $\hat\pi_k > 1/r_k$; otherwise $\hat\xi_k=0$.

For the $\chi^2$ model they are
%
%
\begin{equation}
\label{eq:chisq}
\hat\xi^2 = \frac{ (S_k^2+ Y_k^2 + 3 )}{Y_k-1},\quad
\hat\pi_k = \frac{Y_k-1}{\hat\xi_k^2},
\end{equation}
provided $Y_k>1$ and $1/r_k < \hat\pi_k < (r_k-1)/r_k$; otherwise
$\hat\xi_k=0$.

\item[5.] For each of the $k$ groups, construct weights $w(\hat\xi_k)$.
It is apparent in Figure~\ref{fig::linkeg} that if $|\hat\xi_k|< \delta$, for~$\delta$ near 0, then $w(\hat\xi_k)\approx0$ and it is unlikely
that any tests in the $k$th group will be significant, regardless
of the $p$-value. The stochastic quantity $\delta$ depends upon the relative
values of $(\hat\xi_1,\ldots,\hat\xi_K)$, and the number of
elements in each
group. For this reason we have found that smoothing the weights
generally improves the power of the procedure. We suggest using a linear
combination such as
\[
\hat w_k = (1-\gamma) w(\hat\xi_k) +
\gamma K^{-1}\sum_k w(\hat\xi_k),
\]
with $\gamma=$ 0.01 or 0.05.
The larger the choice of $\gamma$, the more evenly distributed
the weights across groups. Alternatively, one
could smooth the weights by
using a Stein shrinkage estimator or bagging procedure to obtain a
more robust estimator of $(\xi_1,\ldots,\xi_K)$ (Hastie, Tibshirani
and Friedman, \citeyear{hastieetal2001}).
Regardless of how the weights are smoothed, one should renorm them to
ensure the weights sum to $m$. Each test in group $k$ receives the
weight $\hat w_k$. Another effect of the smoothing is to ensure that
each group gets a weight greater than~0.
\end{longlist}

This weighting scheme relies on data-based estimators of the optimal
weights, but with a partition \mbox{of the} data sufficiently crude to
preserve the control of family-wise error rate. The approach is an
example of the ``sieve principle'' (Bickel et~al., \citeyear
{bickeletal1993}). The sieve
principle works because the number of parameters estimated is far less
than the number of observations. Thus, many observations are used to
estimate each parameter. Consequently, parameters are estimated with
substantially less variability than if they were estimated using only
the test statistics from the particular gene under investigation.
Because the weights are determined by the size of the tests in the
entire cluster, the probability of upweighting simply because a single
test is large, due to chance, is small.

\section{Examples}\label{sec6}

\subsection{Binary Weights}
In a study of nicotine dependence, Saccone et~al. (\citeyear{sacconeetal2007}) used binary
weights in a candidate gene study. Their study involved 3713 genetic
variants (single nucleotide polymorphisms or SNPs)
encompassing 348 genes. The genes were divided into two types: 52
nicoinic and dopaminergic receptor genes; and 296 other candidate
genes. Each SNP associated with a gene in the first group was
allocated ten times the weight of a gene in the other category. Using
a generous false discovery rate ($\alpha=0.4$), they identified 39
SNPs; 78\% of these were nicotine receptors, in
contrast to the fraction of nicotine receptors overall (15\%).

\subsection{Independent Data Weights}
For family-based study designs, tests of association are based on
transmission data. In these studies, data are available from which
one can compute the potential power to detect a signal at each SNP
tested; see Ionita-Laza et~al. (\citeyear{ionitalaza2007}) for a
detailed explanation of
this unique feature of family-based data. Because the data used to
calculate the power are independent of the test statistics for
association, these data are available for construction of the weights.
Motivated by this possibility, Ionita-Laza et~al. (\citeyear
{ionitalaza2007}) developed a
weighting scheme. Using independent data, they ranked the SNPs
from most to least promising, in terms of power. They then
constructed an exponential weighting scheme, based on simulations of
genetic models. The scheme results in a small number of SNPs
receiving a top weight, successively more SNPs receiving
correspondingly lower weights, and finally a large number receiving
the lowest weight. In their simulations they found that the power of
the test can often be doubled using this procedure. Using the FHS
data, they apply the technique to a genome-wide association study with
116,204 SNPs and 923 participants. The phenotype of interest is
height. Using their weighting scheme, they obtained one significant
result with weights and none without weights.

%
\begin{figure*}

\includegraphics{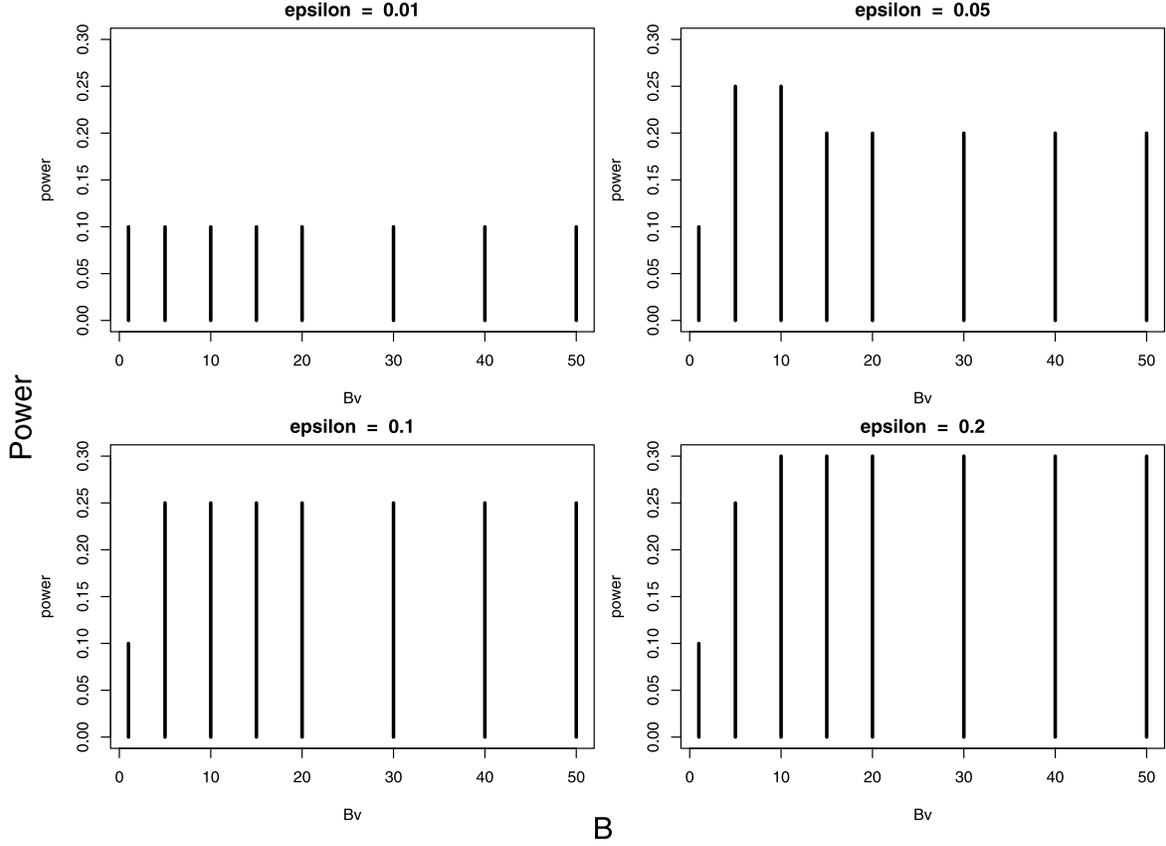}

\caption{Power as a function of B and $\varepsilon$.}\label{fig::linkpow}
\end{figure*}

\subsection{Linkage Weights}

Finding variation in the genetic code that increases the risk for
complex diseases, such as Type II diabetes and schizophrenia, is
critically important to the advancement of genetic epidemiology. In
the\break \hyperref[sec1]{Introduction} we describe a means by which weights could be
extracted from linkage data. Here we illustrate the idea with both
data and simulations.

In the analysis of 955 cases and 1498 controls enrolled in a
genome-wide association study, McQueen and colleagues (\citeyear
{mcqueenetal2008}) used
weights derived from published linkage results. They combined results
from 11
linkage studies on bipolar disorder to obtain $Z$ scores
corresponding to the locations of each association test. From the
linkage results they computed weighted $p$-values using the cummulative normal
weight function (Roeder et~al., \citeyear{roederetal2006}). Although
none of their results were
genome-wide significant, they obtained promising results in four
regions. Three of these are obtained due to strong $p$-values in
combination with a linkage peak. One signal did not correspond to a
linkage peak, but continued to be in the top tier of $p$-values, after
weights were applied.

To illustrate how binary weights could be derived from such linkage data,
we present a realistic synthetic example. Using the methods described
in\break Roeder et~al. (\citeyear{roederetal2006}), we create a linkage
trace that captures many of the features found in actual linkage
traces. In this simulation we generate a full genome (23 chromosomes)
and place 20 disease variants at random, one per chromosome. The
signals from these variants were designed to yield weak signals with
broad peaks. Next, we simulated 100,000 normally distributed
association test statistics mapped to the same genome. Again, 20 of
these tests were generated under the alternative hypothesis of
association. These signals were also weak.

To illustrate the synthetic data, six typical chromosomes are
displayed in Figure~\ref{fig::linkeg}. Each displayed chromosome has
one true signal, with the association test statistic at that location
indicated by an upspike; none of the association tests generated under
the null hypothesis are plotted. Without weights, only 2 of the 20
signals could be detected using a Bonferonni correction. Using binary
weights, as described above, with $\varepsilon=0.05$ and $B=10$, we
discover 5 of the 20 signals. In the left column of the figure all three
signals were discovered, while in the right column none were
discovered (indicated by presence of a down-spike). Comparing the top
row, we see that both signals were up-weighted in the correct location,
but the association signal was not strong enough in the top right
chromosome to achieve significance. Alternatively, in the bottom left
panel the association statistic was substantial enough to reject the
null hypothesis without the benefit of up-weighting.

To examine the robustness of the procedure to choice of weights, we tried
4 choices of $\varepsilon$ (0.01, 0.05, 0.1, 0.2) with $1 \leq B \leq50$. We
made no false discoveries with any of these choices. The power is
displayed in Figure~\ref{fig::linkpow}. To assist in the choice of
parameters, we have found it helpful to examine the number of
discoveries for each choice. In this example, the number of
discoveries varied between 2 (unweighted, i.e., $B=1$) to 6
($\varepsilon=0.2$, $B \geq10$). Five discoveries were made for a broad
range of choices. In principle, choosing $(\varepsilon,B)$ to maximize the
number of discoveries can inflate the error rate. In our simulations
we have found that searching within the family of weights defined by 1
or 2 parameters, such as this binary weight system based upon a
linkage trace, tends to provide very close to nominal protection
against false discoveries.

\section{Discussion}

Several authors have explored the effect of weights on the power of
multiple testing procedures [e.g., Westfall, Kropf and Finos (\citeyear{westfalletal2004})]. These
investigations show that the power of multiple testing procedures can
be increased by using weighted $p$-values. Here we derive the optimal
weights for a commonly used family of tests and show that the power is
remarkably robust to misspecification of these weights.

The same ideas used here can be applied to other testing methods to
improve power. In particular,\break weights can be added to the FDR method,
Holm's stepdown test, and the Donoho and Jin (\citeyear
{donohojin2004}) method. Weighting
ideas can also be used for confidence intervals. Another open
question is the connection with Bayesian methods which have already
been developed to some extent in Efron et~al. (\citeyear{efronetal2001}).

GWAS for some phenotypes such as Type 1
diabetes have yielded exciting results (Todd et~al., \citeyear{toddetal2007}),
while results for other complex diseases have been much less
successful. Presumably, many studies do not have sufficient power to
detect the genetic variants associated with the phenotypes,
even though thousands of cases and
controls have been genotyped. To bolster power, we recommend
up-weighting and down-weighting hypotheses, based on prior likelihood of
association with the phenotype. For instance, Wang, Li and Bucan
(\citeyear{wanglibucan2007})
describe pathway-based approaches for the analysis of GWAS.

Multiple testing arises in GWAS analyses in other contexts as well.
Frequently, multiple tests, assuming different genetic models, are
applied to each genetic marker. Multiple markers in a neighborhood
can be analyzed simultaneously to increase the signal, using
haplotypes, multivariate models and fine-mapping techniques. Data
are often collected in multiple stages of the experiment, and at each
stage promising markers are tested for association. In summary, many
questions concerning multiple testing remain open in the context of
GWAS.

\begin{appendix}
\renewcommand{\theequation}{\arabic{equation}}
\section*{Appendix}\label{appendix}

\begin{pf*}{Proof of Lemma~\ref{lemma::basic}}
The familywise error is
\begin{eqnarray*}
\mathbb{P}\bigl((\mathcal{R}\cap\mathcal{H}_0) > 0\bigr)
& = & \mathbb{P}\biggl( P_j \le\frac{\alpha w_j }{m} \mbox{ for some }j
\in\mathcal{H}_0 \biggr)\\
& \leq& \sum_{j\in\mathcal{H}_0} \mathbb{P}\biggl( P_j \le\frac{\alpha w_j }{m} \biggr)
=\frac{\alpha}{m} \sum_{j\in\mathcal{H}_0} w_j
\\
&\leq&\alpha\overline {w} = \alpha.
\end{eqnarray*}
\upqed
\end{pf*}

\begin{pf*}{Proof of Lemma~\ref{lemma::basic2}}
The familywise error is
\begin{eqnarray*}
&&\mathbb{P}\bigl((\mathcal{R}\cap\mathcal{H}_0) > 0\bigr)
\\
&&\quad =  \mathbb{P}\biggl( P_j \le\frac{\alpha W_j }{m} \mbox{ for some }j\in\mathcal{H}_0 \biggr)\\
&&\quad \leq \sum_{j\in\mathcal{H}_0} \mathbb{P}\biggl( P_j \le\frac{\alpha W_j }{m}\biggr)
\\
&&\quad=\sum_{j\in\mathcal{H}_0} \mathbb{E}_H \biggl(\mathbb{P} \biggl( P_j \le\frac
{\alpha w_j }{m}\Bigm| W_j = w_j \biggr) \biggr)\\
&&\quad= \sum_{j\in\mathcal{H}_0} \mathbb{E}_H(\alpha W_j/m)
= \frac{\alpha}{m} \sum_{j\in\mathcal{H}_0} \mathbb{E}_H(W_j)\\
&&\quad \leq \frac{m_0 \alpha}{m} \leq\alpha.
\end{eqnarray*}
\upqed
\end{pf*}

\begin{pf*}{Proof of Theorem~\ref{thm::optimal}}
Let $C$ denote the set of hypotheses with $\xi_j > 0$. Power is optimized
if $w_j = 0$ for $j \notin C$. The average power is
\[
\frac1 {m_1} \sum_{j\in C}
\phib\biggl(\phibi\biggl(\frac{\alpha w_j}m \biggr)-\xi_j \biggr),
\]
with constraint
\[
\sum w_j = m.
\]
Choose $\underline w$ to maximize
\begin{eqnarray*}
\pi=\frac1 {m_1} \sum_{j \in C} \phib\biggl(\phibi\biggl(\frac{\alpha w_j}m \biggr)- \xi_j \biggr)
- \lambda\Bigl({m-\sum w_i} \Bigr)\\
\end{eqnarray*}
by setting the derivative to zero
\begin{eqnarray*}
\frac\partial{\partial w_i}\pi&=&
- \lambda+\frac{\phi(\phibi(\alpha w_j/m ) - \xi_j )}
{\phi(\phibi(\alpha w_j/m ) )} \frac{\alpha}m = 0,
\\
\frac{m\lambda}{\alpha} &=& \frac{\phi(\phibi(\alpha w_j/m) - \xi_j )}{\phi(\phibi(\alpha w_j/m ) )}.
\end{eqnarray*}
The $\underline w$ that solves these equations is given in (\ref{eq:wq1}).
Finally, solve for $c$ such that $\sum_i w_i = m$.
\end{pf*}

\begin{pf*}{Proof of Theorem~\ref{thm::sparse-is-good}}
The first statement follows easily by noting that the worst case
corresponds to choosing weight $B$ in the first term in $R(\xi)$ and
choosing weight $b$ in the second term in $R(\xi)$.
The rest follows by Taylor expanding
$R_{b,B}(\xi)$ around $b=1$.
\end{pf*}

\begin{pf*}{Proof of Lemma~\ref{lemma::safe}}
With $b=0$, $R_{b,B}(\xi) \geq0$ when
%
%
\begin{equation}\label{eq::bis0}
\overline{\Phi}(z_{B \alpha/m} - \xi) -2 \overline{\Phi
}(z_{\alpha/m} - \xi) \geq0.
\end{equation}
With $B\geq2$, (\ref{eq::bis0}) holds at $\xi=0$.
The left-hand side is increasing in $\xi$ for $\xi$ near 0,
but (\ref{eq::bis0}) does not hold at $\xi= z_{\alpha/m}$.
So (\ref{eq::bis0}) must hold in the interval
$[0,\xi_*]$.
Rewrite (\ref{eq::bis0}) as
$\overline{\Phi}(z_{B \alpha/m} - \xi) - \overline{\Phi
}(z_{\alpha/m} - \xi) \geq\overline{\Phi}(z_{\alpha/m} - \xi)$.
We lower bound the left-hand side and upper bound the right-hand side.
The left-hand side is
$\overline{\Phi}(z_{B \alpha/m} - \xi) - \overline{\Phi
}(z_{\alpha/m} - \xi) =
\int_{ z_{B \alpha/m} - \xi}^{ z_{\alpha/m} - \xi} \phi(u)\, du
\geq
(z_{\alpha/m} - z_{B\alpha/m})\phi(z_{\alpha/m}-\xi)$.
The right-hand side
can be bounded using Mill's ratio:
$\overline{\Phi}(z_{\alpha/m} - \xi) \leq\break\phi(z_{\alpha/m} - \xi
)/(z_{\alpha/m} - \xi)$.
Set the lower bound greater than the upper bound to obtain the stated result.
\end{pf*}

\begin{pf*}{Proof of Lemma~\ref{lemma::discon}}
Choose $K> 1$ such that
$1/(K+1) < 1/a - \varepsilon$.
Choose $1> \gamma> (2\alpha-a)/K$.
Choose a small $c>0$.
Let
$\xi= A + \sqrt{A^2 - 2c}$ and
$u = B - \sqrt{B^2 - 2c}$,
where
\begin{eqnarray*}
A = \overline{\Phi}^{-1} \biggl(\frac{\alpha}{(m(\gamma K+a))} \biggr),
\\
B = \overline{\Phi}^{-1} \biggl(\frac{K \alpha}{(m(\gamma K+a))} \biggr).
\end{eqnarray*}
Then $\rho(\xi)=1/a$ and
$\tilde{\rho}(\xi)=1/(K+1)$.
Now $d(Q,\break\tilde{Q})=\gamma$.
Taking $K$ sufficiently large and
$\gamma$ sufficiently close to $(2\alpha-a)/K$
makes $\gamma< \delta$.
\end{pf*}

It is convenient to prove Theorem~\ref{thm::drop-u}
before proving Theorem~\ref{thm::restrict-u}.

\begin{pf*}{Proof of Theorem~\ref{thm::drop-u}}
Let $c_*$ solve
%
%
\begin{equation}\label{eq::define-c-star}
\gamma\overline{\Phi}\bigl(\sqrt{2 c_*}\bigr) + a \overline{\Phi} \biggl( \frac{\xi}{2}+\frac{c_*}{\xi} \biggr)=\frac{\alpha}{m}.
\end{equation}
We claim first that, for any $c>c_*$,
there is no $u$ such that
the weights average to 1.
Fix $c > c_*$.
The weights average to 1 if and only if
%
%
\begin{equation}\label{eq::we-need}
\gamma\overline{\Phi} \biggl(\frac{c}{u}+\frac{u}{2} \biggr) +
a \overline{\Phi} \biggl( \frac{\xi}{2}+\frac{c}{\xi} \biggr)= \frac{\alpha}{m}.
\end{equation}
Since $c > c_*$ and since the second term
is decreasing in~$c$, we must have
\[
\overline{\Phi} \biggl(\frac{c}{u}+\frac{u}{2} \biggr) >
\overline{\Phi}\bigl(\sqrt{2 c_*}\bigr).
\]
The function
$r(u)=\overline{\Phi}(c/u+u/2)$ is maximized at
$u=\sqrt{2c}$.
So $r(\sqrt{2c}) \geq r(u)$.
But $r(\sqrt{2c}) = \overline{\Phi}(\sqrt{2 c})$.
Hence,
$\overline{\Phi}(\sqrt{2 c}) \geq r(u) \geq
\overline{\Phi}(\sqrt{2 c_*})$.
This implies $c < c_*$, which is a contradiction.
This establishes that $\sup_u c(u) \leq c_*$.
On the other hand,
taking
$c=c_*$ and $u=\sqrt{2c_*}$ solves equation
(\ref{eq::we-need}). Thus, $c_*$ is indeed the largest $c$ that solves the
equation which establishes the first claim.
The second claim follows by noting that
\[
\gamma\overline{\Phi}\bigl(\sqrt{2 c_*}\bigr) + a \overline{\Phi}
\biggl( \frac{\xi}{2}+\frac{c_*}{\xi} \biggr)=
\gamma\overline{\Phi}\bigl(\sqrt{2 c_*}\bigr) + O(a).
\]
Now set this expression equal to $\alpha/m$ and solve.
\end{pf*}

\begin{pf*}{Proof of Theorem~\ref{thm::restrict-u}}
Define $c_*$ as in (\ref{eq::define-c-star}).
If $u_*=\sqrt{2c_*} \leq\xi$, then the
the proof proceeds as in the previous proof.
So we first need to establish for which values of $\xi$
is this true.
Let $r(c) = \gamma\overline{\Phi}(\sqrt{2c}) + a\overline{\Phi
}(\xi/2 + c/\xi)$.
We want to find out when
the solution of $r(c)=\alpha/m$
is such that $\sqrt{2c} \leq\xi$, or,
equivalently, $c \leq\xi^2/2$.
Now $r$ is decreasing in $c$.
Since $\gamma+ a \geq\alpha/m$,
$r(-\infty) \geq\alpha/m$.
Hence, there is a solution
with $c \leq\xi^2/2$
if and only if
$r(\xi^2/2) \leq\alpha/m$.
But
$r(\xi^2/2) = (\gamma+ a)\overline{\Phi}(\xi),$
so we conclude that
there is such a solution if and only if
$(\gamma+ a)\overline{\Phi}(\xi) \leq\alpha/m$, that is,
$\xi\geq z_{\alpha/(m(\gamma+a))}=\xi_0$.

Now suppose that
$\xi< \xi_0$.
We need to find $u\leq\xi$ to make $c$ as large as possible
in the equation
$v(u,c)\equiv\gamma\overline{\Phi}(u/2 + c/u) + a \overline{\Phi
}(\xi/2 + c/\xi)=\alpha/m$.
Let $u_*=\xi$ and $c_*=\xi z_{\alpha/(m(\gamma+a))}-\xi^2/2$.
By direct substitution,
$v(u_*,c_*)=\alpha/m$ for this choice of $u$ and $c$
and, clearly, $u_*\leq\xi$ as required.
We claim that this is the largest possible $c_*$.
To see this, note that
$v(u,c)< v(u,c_*)$.
For $\xi\leq\xi_0$,
$v(u,c_*)$ is a decreasing function of $u$.
Hence,
$v(u,c) < v(u,c_*) \leq v(u_*,c_*)=\alpha/m$.
This contradicts the fact that
$v(u,c) =\alpha/m$.

For the second claim, note that the power
of the weighted test beats the power of Bonferroni
if and only if
the weight $w = (m/\alpha)\overline{\Phi}(\xi/2 + C(\xi)/2) \geq1$,
which is equivalent to
%
%
\begin{equation}\label{eq::this-must}
C(\xi) \leq\xi z_{\alpha/m} - \xi^2/2.
\end{equation}
When $\xi\leq\xi_0$,
$C(\xi) =\xi\xi_0 - \xi^2/2$.
By assumption, $\gamma+ a \leq1$ so that
$z_{\alpha/(m(\gamma+a))}\leq z_{\alpha/m}$ and
now suppose that
$\xi_0 < \xi\leq\xi_*$.
Then $C(\xi)$ is the solution to
$r(c) = \gamma\overline{\Phi}(\sqrt{2c}) + a\overline{\Phi}(\xi
/2 + c/\xi)=\alpha/m$.
We claim that
(\ref{eq::this-must}) still holds.
Suppose not.
Then, since $r(c)$ is decreasing in $c$,
$r(\xi z_{\alpha/m} - \xi^2/2) > r(C(\xi))=\alpha/m$.
But, by direct calculation,
$r(\xi z_{\alpha/m} - \xi^2/2) > \alpha/m$
implies that $\xi> \xi_*$,
which is a contradiction.
Thus, (\ref{eq::thebonfclaim}) holds.

Finally, we turn to (\ref{eq::thenextclaim}).
In this case,
$C(\xi) = z^2_{\alpha/(m\gamma)}/\break 2 + O(a)$.
The worst case power is
$\overline{\Phi}(C(\xi)/\xi- \xi/2)=\overline{\Phi}(z^2_{\alpha
/(m\gamma)}/(2\xi) - \xi/2) + O(a)$.
The latter is increasing in $\xi$ and so is at least
$\overline{\Phi}(z^2_{\alpha/(m\gamma)}/(2\xi_*) - \xi_*/2) +
O(a) =
\overline{\Phi}((z^2_{\alpha/(m\gamma)}/(2\xi_*) - \xi_*^2)/(2\xi
_*))+O(a)$, as\break claimed.
The next two equations follow from standard tail approximations for
Gaussians.
Specifically, a Gaussian quantile
$z_{\beta/m}$ can be written as
$z_{\beta/m}=\sqrt{2\log(m L_m/\beta)}$,
where $L_m = c \log^a (m)$ for constants $a$ and $c$
[Donoho and Jin (\citeyear{donohojin2004})].
Inserting this into the previous expression yields the final expression.
\end{pf*}

\begin{pf*}{Proof of Theorem~\ref{thm::method1}}
Setting $\pi(w,\xi_m) = \overline{\Phi}(\overline{\Phi
}^{-1}(w\alpha/m) - \xi_m)$
equal to $1-\beta$ implies
$w = (m/\alpha)\overline{\Phi}(z_{1-\beta} + z_{\alpha/m})$,
which is equal to $w_1$ as stated in the theorem.
The stated form of $w_0$ implies that the weights average to 1.
The stated solution thus satisfies
the restriction that a fraction $\varepsilon$ have power at least
$1-\beta$.
Increasing the weight of any hypothesis whose weight is $w_0$ necessitates
reducing the weight of another hypothesis.
This either reduces the minimum power or forces a hypothesis with power
$1-\beta$
to fall below $1-\beta$.
Hence, the stated solution does in fact maximize the minimum power.
\end{pf*}

\end{appendix}

\section*{Acknowledgments}

The authors thank Jamie Robins for helping us to clarify several issues.
Research supported in part by National Institute of Mental Health Grant MH057881 and NSF Grant AST 0434343.


\begin{thebibliography}{99}

\bibitem[\protect\citeauthoryear{Benjamini and Hochberg}{1995}]{benjaminihochberg1995}
\textsc{Benjamini}, Y. and \textsc{Hochberg}, Y. (1995).
Controlling the false discovery rate: A practical and powerful approach
to multiple testing.
\textit{J. Roy. Statist. Soc. Ser. B} \textbf{57} 289--300.
\MR{1325392}

\bibitem[\protect\citeauthoryear{Benjamini and
Hochberg}{1997}]{benjaminihochberg1997}
\textsc{Benjamini}, Y. and \textsc{Hochberg}, Y. (1997).
Multiple hypotheses testing with weights.
\textit{Scand. J. Statist.} \textbf{24} 407--418.
\MR{1481424}

\bibitem[\protect\citeauthoryear{Benjamini, Krieger and
Yekutieli}{2006}]{MR2261438}
\textsc{Benjamini}, Y., \textsc{Krieger}, A.~M. and \textsc
{Yekutieli}, D. (2006).
Adaptive linear step-up procedures that control the false discovery rate.
\textit{Biometrika} \textbf{93} 491--507.
\MR{2261438}

\bibitem[\protect\citeauthoryear{Benjamini and Yekutieli}{2001}]{MR1869245}
\textsc{Benjamini}, Y. and \textsc{Yekutieli}, D. (2001).
The control of the false discovery rate in multiple testing under dependency.
\textit{Ann. Statist.} \textbf{29} 1165--1188.
\MR{1869245}

\bibitem[\protect\citeauthoryear{Bickel et~al.}{1993}]{bickeletal1993}
\textsc{Bickel}, P., \textsc{Klaassen}, C., \textsc{Ritov}, Y. and
\textsc{Wellner}, J. (1993).
Efficient and adaptive estimation for semiparametric models.
Technical report, Johns Hopkins Series in the Mathematical Statistics,
Baltimore, Maryand.
\MR{1245941}


\bibitem[\protect\citeauthoryear{Blanchard and Roquain}{2008}]{MR2448601}
\textsc{Blanchard}, G. and \textsc{Roquain}, E. (2008).
Two simple sufficient conditions for {FDR control.}
\textit{Electron. J. Stat.} \textbf{2} 963--992.
\MR{2448601}

\bibitem[\protect\citeauthoryear{Blanchard and Roquain}{2009}]{blanchardroquain2008b}
\textsc{Blanchard}, G. and \textsc{Roquain}, E. (2009).
Adaptive FDR control under independence and dependence.
\textit{J. Mach. Learn. Res.}
To appear.

\bibitem[\protect\citeauthoryear{Chen et~al.}{2000}]{chenetal2000}
\textsc{Chen}, J.~J., \textsc{Lin}, K.~K., \textsc{Huque}, M. and
\textsc{Arani}, R.~B. (2000).
Weighted $p$-value adjustments for animal carcinogenicity trend test.
\textit{Biometrics} \textbf{56} 586--592.

\bibitem[\protect\citeauthoryear{Donoho and Jin}{2004}]{donohojin2004}
\textsc{Donoho}, D. and \textsc{Jin}, J. (2004).
Higher criticism for detecting sparse heterogeneous mixtures.
\textit{Ann. Statist.} \textbf{32} 962--994.
\MR{2065195}

\bibitem[\protect\citeauthoryear{Efron}{2007}]{efron2007}
\textsc{Efron}, B. (2007).
Simultaneous inference: When should hypothesis testing problems be combined?
\textit{Ann. Appl. Statist.} \textbf{2} 197--223.
\MR{2415600}

\bibitem[\protect\citeauthoryear{Efron et~al.}{2001}]{efronetal2001}
\textsc{Efron}, B., \textsc{Tibshirani}, R., \textsc{Storey}, J.~D.
and \textsc{Tusher}, V. (2001).
Empirical {Bayes analysis of a microarray experiment.}
\textit{J. Amer. Statist. Assoc.} \textbf{96} 1151--1160.
\MR{1946571}

\bibitem[\protect\citeauthoryear{Genovese and
Wasserman}{2002}]{genovesewasserman2002}
\textsc{Genovese}, C. and \textsc{Wasserman}, L. (2002).
Operating characteristics and extensions of the false discovery rate procedure.
\textit{J.~R. Stat. Soc. Ser. B~Stat. Methodol.} \textbf{64} 499--517.
\MR{1924303}

\bibitem[\protect\citeauthoryear{Genovese, Roeder and
Wasserman}{2006}]{grw2006}
\textsc{Genovese}, C.~R., \textsc{Roeder}, K. and \textsc
{Wasserman}, L. (2006).
False discovery control with {$p$-value weighting.}
\textit{Biometrika} \textbf{93} 509--524.
\MR{2261439}

\bibitem[\protect\citeauthoryear{Hastie, Tibshirani and
Friedman}{2001}]{hastieetal2001}
\textsc{Hastie}, T., \textsc{Tibshirani}, R. and \textsc{Friedman},
J. (2001).
\textit{The Elements of Statistical Learning: Data Mining, Inference,
and Prediction}.
Springer, New York.
\MR{1851606}

\bibitem[\protect\citeauthoryear{Holm}{1979}]{holm1979}
\textsc{Holm}, S. (1979).
A simple sequentially rejective multiple test procedure.
\textit{Scand. J.~Statist.} \textbf{6} 65--70.
\MR{0538597}

\bibitem[\protect\citeauthoryear{Ionita-Laza et~al.}{2007}]{ionitalaza2007}
\textsc{Ionita}-\textsc{Laza}, I., \textsc{McQueen}, M., \textsc
{Laird}, N. and \textsc{Lange}, C. (2007).
Genomewide weighted hypothesis testing in family-based association
studies, with an application to a 100K scan.
\textit{Am. J. Hum. Genet.} \textbf{81} 607--614.

\bibitem[\protect\citeauthoryear{Kropf et~al.}{2004}]{kropfetal2004}
\textsc{Kropf}, S., \textsc{L{\"{a}}uter}, J., \textsc{Eszlinger}, M., \textsc
{Krohn}, K. and \textsc{Paschke}, R. (2004).
Nonparametric multiple test procedures with data-driven order of
hypotheses and with weighted hypotheses.
\textit{J.~Statist. Plann. Inference} \textbf{125} 31--47.
\MR{2086887}

\bibitem[\protect\citeauthoryear{McQueen and
colleagues}{2008}]{mcqueenetal2008}
\textsc{McQueen}, M. and colleagues (2008).
Personal communication.

\bibitem[\protect\citeauthoryear{Roeder et~al.}{2006}]{roederetal2006}
\textsc{Roeder}, K., \textsc{Bacanu}, S.-A., \textsc{Wasserman}, L.
and \textsc{Devlin}, B. (2006).
Using linkage genome scans to improve power of association in genome scans.
\textit{Am. J. Hum. Genet.} \textbf{78} 243--252.

\bibitem[\protect\citeauthoryear{Roeder, Wasserman and
Devlin}{2007}]{roederetal2007}
\textsc{Roeder}, K., \textsc{Wasserman}, L. and \textsc{Devlin}, B. (2007).
Improving power in genome-wide association studies: Weights tip the scale.
\textit{Genet. Epidemiol.} \textbf{31} 741--747.

\bibitem[\protect\citeauthoryear{Romano, Shaikh and Wolf}{2008}]{MR2470085}
\textsc{Romano}, J.~P., \textsc{Shaikh}, A.~M. and \textsc{Wolf}, M. (2008).
Control of the false discovery rate under dependence using the
bootstrap and subsampling.
\textit{TEST} \textbf{17} 417--442.
\MR{2470085}

\bibitem[\protect\citeauthoryear{Roquain and van~de
Wiel}{2008}]{roquainvandewiel2008}
\textsc{Roquain}, E. and \textsc{van~de Wiel}, M. (2008).
Multi-weighting for {FDR control.}
Available at
\href{http://arxiv.org/abs/0807.4081}{arXiv:0807.4081}.

\bibitem[\protect\citeauthoryear{Rosenthal and
Rubin}{1983}]{rosenthalrubin1983}
\textsc{Rosenthal}, R. and \textsc{Rubin}, D. (1983).
Ensemble-adjusted $p$-values.
\textit{Psychol. Bull.} \textbf{94} 540--541.

\bibitem[\protect\citeauthoryear{Rubin, Dudoit and van~der
Laan}{2006}]{MR2240850}
\textsc{Rubin}, D., \textsc{Dudoit}, S. and \textsc{van~der Laan},
M. (2006).
A method to increase the power of multiple testing procedures through
sample splitting.
\textit{Stat. Appl. Genet. Mol. Biol.} \textbf{5}, Art. 19 (electronic).
\MR{2240850}

\bibitem[\protect\citeauthoryear{Sabatti, Service and
Freimer}{2003}]{sabattietal2003}
\textsc{Sabatti}, C., \textsc{Service}, S. and \textsc{Freimer}, N. (2003).
False discovery rate in linkage and association genome screens for
complex disorders.
\textit{Genetics} \textbf{164} 829--833.

\bibitem[\protect\citeauthoryear{Saccone et~al.}{2007}]{sacconeetal2007}
\textsc{Saccone}, S., \textsc{Hinrichs}, A.~L.,  \textsc{Saccone}, N.,
\textsc{Chase}, G., \textsc{Konvicka}, K., \textsc{Madden}, P.,
\textsc{Breslau}, N., \textsc{Johnson}, E., \textsc{Hatsukami}, D.,
\textsc{Pomerleau}, O., \textsc{Swan}, G., \textsc{Goate}, A.,
\textsc{Rutter}, J., \textsc{Bertelsen}, S., \textsc{Fox}, L.,
\textsc{Fugman}, D., \textsc{Martin}, N., \textsc{Montgomery}, G.,
\textsc{Wang}, J., \textsc{Ballinger}, D., \textsc{Rice}, J. and
\textsc{Bierut}, L. (2007).
Cholinergic nicotinic receptor genes implicated in a nicotine
dependence association study targeting 348 candidate genes with 3713 SNPs.
\textit{Hum. Mol. Genet.} \textbf{16} 36--49.

\bibitem[\protect\citeauthoryear{Sarkar and Heller}{2008}]{sarkarheller2008}
\textsc{Sarkar}, S. and \textsc{Heller}, R. (2008).
Comments on: Control of the false discovery rate under dependence using
the bootstrap and subsampling.
\textit{TEST} \textbf{17} 450--455.
\MR{2470085}

\bibitem[\protect\citeauthoryear{Sarkar}{2002}]{MR1892663}
\textsc{Sarkar}, S.~K. (2002).
Some results on false discovery rate in stepwise multiple testing procedures.
\textit{Ann. Statist.} \textbf{30} 239--257.
\MR{1892663}

\bibitem[\protect\citeauthoryear{Satagopan and
Elston}{2003}]{satagopanelston2003}
\textsc{Satagopan}, J. and \textsc{Elston}, R. (2003).
Optimal two-stage genotyping in population-based association studies.
\textit{Genet. Epidemiol.} \textbf{25} 149--157.

\bibitem[\protect\citeauthoryear{Schuster, Kropf and
Roeder}{2004}]{schusteretal2004}
\textsc{Schuster}, E., \textsc{Kropf}, S. and \textsc{Roeder}, I. (2004).
Micro array based gene expression analysis using parametric
multivariate tests per gene---a generalized application of multiple
procedures with data-driven order of hypotheses.
\textit{Biom. J.} \textbf{46} 687--698.
\MR{2108612}

\bibitem[\protect\citeauthoryear{Signoravitch}{2006}]{signoravitch2006}
\textsc{Signoravitch}, J. (2006).
Optimal multiple testing under the general linear model.
Technical report, Harvard Biostatistics.

\bibitem[\protect\citeauthoryear{Skol et~al.}{2006}]{skoletal2006}
\textsc{Skol}, A., \textsc{Scott}, L., \textsc{Abecasis}, G. and
\textsc{Boehnke}, M. (2006).
Joint analysis is more efficient than replication-based analysis for
two-stage genome-wide association studies.
\textit{Nat. Genet.} \textbf{38} 390--394.

\bibitem[\protect\citeauthoryear{Spj{\o}tvoll}{1972}]{spjotvoll1972}
\textsc{Spj{\o}tvoll}, E. (1972).
On the optimality of some multiple comparison procedures.
\textit{Ann. Math. Statist.} \textbf{43} 398--411.
\MR{0301871}

\bibitem[\protect\citeauthoryear{Storey}{2002}]{storey2002}
\textsc{Storey}, J.~D. (2002).
A direct approach to false discovery rates.
\textit{J. R. Stat. Soc. Ser. B Stat. Methodol.} \textbf{64} 479--498.
\MR{1924302}


\bibitem[\protect\citeauthoryear{Storey}{2007}]{storey2007}
\textsc{Storey}, J.~D. (2007).
The optimal discovery procedure: A new approach to simultaneous
significance testing.
\textit{J. R. Stat. Soc. Ser. B Stat. Methodol.} \textbf{69} 347--368.
\MR{2323757}

\bibitem[\protect\citeauthoryear{Storey and
Tibshirani}{2003}]{storeytibshirani2003}
\textsc{Storey}, J. and \textsc{Tibshirani}, R. (2003).
Statistical significance for genome-wide studies.
\textit{Proc. Natl. Acad. Sci. USA} \textbf{100} 9440--9445.
\MR{1994856}

\bibitem[\protect\citeauthoryear{Sun and Cai}{2007}]{suncai2007}
\textsc{Sun}, W. and \textsc{Cai}, T.~T. (2007).
Oracle and adaptive compound decision rules for false discovery rate control.
\textit{J. Amer. Statist. Assoc.} \textbf{102} 901--912.
\MR{2411657}

\bibitem[\protect\citeauthoryear{Todd et~al.}{2007}]{toddetal2007}
\textsc{Todd}, J., \textsc{Walker}, N., \textsc{Cooper}, J.,
\textsc{Smyth}, D., \textsc{Downes}, K., \textsc{Plagnol}, V., \textsc{Bailey}, R.,
\textsc{Nejentsev}, S., \textsc{Field}, S., \textsc{Payne}, F.,
\textsc{Lowe}, C., \textsc{Szeszko}, J., \textsc{Hafler}, J.,
\textsc{Zeitels}, L., \textsc{Yang}, J., \textsc{Vella}, A., \textsc
{Nutland}, S., \textsc{Stevens}, H., \textsc{Schuilenburg}, H.,
\textsc{Coleman}, G., \textsc{Maisuria}, M., \textsc{Meadows}, W.,
\textsc{Smink}, L.~J., \textsc{Healy}, B., \textsc{Burren}, O., \textsc{Lam},
A., \textsc{Ovington}, N., \textsc{Allen}, J., \textsc{Adlem}, E.,
\textsc{Leung}, H., \textsc{Wallace}, C., \textsc{Howson}, J.,
\textsc{Guja}, C., \textsc{Ionescu}-\textsc{Tirgovi}, C., \textsc{Genetics of Type~1 Diabetes in Finland},
\textsc{Simmonds}, M., \textsc{Heward}, J., \textsc{Gough}, S.,\
\textsc{Wellcome Trust Case Control
Consortium}, \textsc{Dunger}, D., \textsc{Wicker}, L. and \textsc{Clayton}, D. (2007).
Robust associations of four new chromosome regions from genome-wide
analyses of type 1 diabetes.
\textit{Nat. Genet.} \textbf{39} 857--864.

\bibitem[\protect\citeauthoryear{Wang et~al.}{2006}]{wangetal2006}
\textsc{Wang}, H., \textsc{Thomas}, D., \textsc{Pe'er}, I. and
\textsc{Stram}, D. (2006).
Optimal two-stage genotyping designs for genome-wide association scans.
\textit{Genet. Epidemiol.} \textbf{30} 356--368.

\bibitem[\protect\citeauthoryear{Wang, Li and Bucan}{2007}]{wanglibucan2007}
\textsc{Wang}, K., \textsc{Li}, M. and \textsc{Bucan}, M. (2007).
Pathway-based approaches for analysis of genomewide association studies.
\textit{Am. J. Hum. Genet.} \textbf{81} 1278--1283.

\bibitem[\protect\citeauthoryear{Westfall, Krishen and
Young}{1998}]{westfallkrishenyoung1998}
\textsc{Westfall}, P., \textsc{Krishen}, A. and \textsc{Young}, S. (1998).
Using prior information to allocate significance levels for multiple endpoints.
\textit{Stat. Med.} \textbf{17} 2107--2119.

\bibitem[\protect\citeauthoryear{Westfall and
Krishen}{2001}]{westfallkrishen2001}
\textsc{Westfall}, P.~H. and \textsc{Krishen}, A. (2001).
Optimally weighted, fixed sequence and gatekeeper multiple testing procedures.
\textit{J.~Statist. Plann. Inference} \textbf{99} 25--40.
\MR{1858708}

\bibitem[\protect\citeauthoryear{Westfall, Kropf and Finos}{2004}]{westfalletal2004}
\textsc{Westfall}, P.~H., \textsc{Kropf}, S. and \textsc{Finos}, L. (2004).
Weighted {FWE-controlling methods in high-dimensional situations.}
In \textit{Recent Developments in Multiple Comparison Procedures}.
\textit{IMS Lecture Notes Monogr. Ser.} \textbf{47} 143--154.
IMS, Beachwood, OH.
\MR{2118598}

\bibitem[\protect\citeauthoryear{Westfall and
Soper}{2001}]{westfallsoper2001}
\textsc{Westfall}, P.~H. and \textsc{Soper}, K.~A. (2001).
Using priors to improve multiple animal carcinogenicity tests.
\textit{J. Amer. Statist. Assoc.} \textbf{96} 827--834.
\MR{1963409}
\end{thebibliography}
\end{document}